\documentclass[12pt,preprint]{aastex}  

\usepackage{fancybox}
\usepackage{epsfig}
\usepackage{epsf}
\usepackage{colordvi}
\usepackage[usenames,dvipsnames]{color}

\usepackage{ulem}
\usepackage{natbib}
\usepackage{graphicx}
\usepackage{txfonts}
\newcommand{\va}{v_{\rm A}}
\newcommand{\omegaA}{\omega_{\rm A}}
\newcommand{\vs}{v_{\rm s}}
\newcommand{\omegaC}{\omega_{\rm c}}
\newcommand{\omegaf}{\omega_{\rm f}}
\newcommand{\omegasl}{\omega_{\rm sl}}
\newcommand{\ra}{r_{\rm A}}
\newcommand{\rc}{r_{\rm c}}
\newcommand{\rhoi}{\rho_{\rm i}}
\newcommand{\rhoe}{\rho_{\rm e}}
\newcommand{\vai}{v_{\rm A,i}}
\newcommand{\vae}{v_{\rm A,e}}
\newcommand{\omegaAi}{\omega_{\rm A,i}}
\newcommand{\omegaAe}{\omega_{\rm A,e}}

\newcommand{\vk}{v_{\rm k}}
\newcommand{\xii}{\mbox{\boldmath{$\xi$}}}

\begin{document}
\title{Surface Alfv\'en waves in solar flux tubes}
\shorttitle{Surface Alfv\'en waves}
\shortauthors{Goossens et al.}
\author{M. Goossens$^1$,  J. Andries$^1$, R. Soler$^1$, T. Van Doorsselaere$^{1,3}$, I. Arregui$^2$, \& J. Terradas$^2$}
\affil{$^1$Centre for Plasma Astrophysics, Department of Mathematics, Katholieke Universiteit Leuven,
              Celestijnenlaan 200B, 3001 Leuven, Belgium}
              \email{marcel.goossens@wis.kuleuven.be}  
\affil{$^2$Solar Physics Group, Departament de F\'isica, Universitat de les Illes Balears,
              E-07122, Palma de Mallorca, Spain}
\affil{$^3$Postdoctoral Fellow of the FWO Vlaanderen}

\begin{abstract}
Magnetohydrodynamic (MHD) waves are ubiquitous in the solar atmosphere. Alfv\'{e}n waves and magneto-sonic waves are  particular classes of MHD waves. These wave modes are clearly different and have pure properties  in uniform plasmas of infinite extent only. Due to plasma non-uniformity MHD waves have mixed properties and cannot be classified as pure Alfv\'{e}n or magneto-sonic waves. However, vorticity is a quantity unequivocally related to Alfv\'en waves as compression is for magneto-sonic waves. Here, we investigate MHD waves superimposed on a one-dimensional non-uniform straight cylinder with constant magnetic field.  For a piecewise constant density profile we find that the fundamental radial modes of the non-axisymmetric waves have the same properties as surface Alfv\'en waves at a true discontinuity in density. Contrary to the classic Alfv\'en waves in a uniform plasma of infinite extent, vorticity is zero everywhere except at the cylinder boundary. If the discontinuity in density is replaced with a continuous variation of density, vorticity is spread out over the whole interval with non-uniform density. The fundamental radial modes of the non-axisymmetric waves do not need compression to exist unlike the radial overtones. In thin magnetic cylinders the fundamental radial modes of the non-axisymmetric waves with phase velocities between the internal and the external Alfv\'en velocities can be considered as surface Alfv\'{e}n waves. On the contrary, the radial overtones can be related to fast-like magneto-sonic modes.

\end{abstract}

\keywords{Sun: oscillations --- Sun: corona --- Sun: atmosphere --- Magnetohydrodynamics (MHD) --- Waves}
       
%\maketitle

%\begin{tom}Tom's additions will be in this nice colour.\end{tom}
%\begin{rob}Roberto's additions will be in this (also) nice colour.\end{rob}
%\begin{jes}Jesse's additions will be in this (also) nice colour.\end{jes}

\section{Introduction}
Alfv\'{e}n waves were predicted  70 years ago by H. Alfv\'{e}n in a short paper entitled ``Existence of Electromagnetic-Hydrodynamic Waves'' \citep{alfven1942}.  Alfv\'{e}n's  prediction was initially met with disbelief and was accepted only years later \citep{faldes06}. Alfv\'{e}n waves are ubiquitous in magnetized plasmas, in fusion plasma physics, geophysics, astrophysics and solar physics \citep[see recent review by][]{gekelman11}.

Alfv\'{e}n waves are a particular class of magnetohydrodynamic (MHD) waves.  MHD waves  have become a subject of intense research in solar physics, largely because observations clearly show that  they are ubiquitous in the solar atmosphere \citep[e.g.][]{tomczyk2007,depontieu2007,mcintosh2011}. A pioneering theory paper on MHD waves is the  paper by \citet{edwin1983} on  MHD waves on an axisymmetric magnetic cylinder of piecewise constant  density and constant straight magnetic field \citep[see also, e.g.,][]{wentzel1979,spruit1982}. The paper by \citet{edwin1983} has paved the  way for new research as observations brought to light new information on MHD waves  that was unknown in the 1980s. For MHD waves in the solar corona Figure 4 of \citet{edwin1983} is often used as reference frame.  In this figure a variety of MHD waves are displayed. Fast and slow waves are present but Alfv\'{e}n waves are apparently absent from the diagram. The question then arises, where are the Alfv\'{e}n waves? 

Compression and vorticity are key quantities for characterizing MHD waves. In a uniform plasma of infinite extent MHD waves have non-zero compression and zero vorticity in case of magneto-sonic waves, or zero compression and non-zero vorticity in case of Alfv\'en waves. The basic characteristics of the classic Alfv\'{e}n wave are that its motions are vortical and the total pressure in the plasma remains constant during the passage of the wave. For an inhomogeneous medium, however, the total pressure, in general, couples with the dynamics of the motion, and the assumption of neglect of pressure perturbations becomes invalid \citep{hasegawauberoi}.  However, that does not mean that the concept of Alfv\'{e}n waves is obsolete. In general in an inhomogeneous plasma, MHD waves have mixed properties which can be traced back to the properties of the classic slow, fast, and Alfv\'{e}n waves in a homogeneous plasma of infinite extent. The degree to which the classic properties are present in a given MHD wave depends on the background through which the MHD wave propagates. The phenomenon of MHD waves with mixed properties or coupled waves {\bf can lead to damping} and was discussed by, e.g., \citet{chen-hasegawa1974b,tirry1996,goossens2001,goossens2002,goossens2002sant,goossens2011,degroof2002,terradas2008,callygoossens2008,pascoe2010,pascoe2011,cally-andries2010}, among many others. 

An MHD wave on an axi-symmetric one-dimensional cylindrical plasma is characterized by two wavenumbers, i.e., the azimuthal wavenumber, $m$, and the axial wavenumber, $k_z$. In addition, modes can have different nodes in the radial direction and this number of nodes can be used to further classify the modes. The term fundamental radial mode used here refers to waves that have no nodes in the radial part of the eigenfunction. The main objective of this paper is to show that the fundamental radial modes of the non-axisymmetric MHD waves with phase velocities between the internal, $\vai$, and the external, $\vae$, Alfv\'en velocities are surface Alfv\'en waves. These modes were originally called fast waves by \citet{edwin1983}, and the adjective `fast' became widely used in papers that followed the nomenclature of \citet{edwin1983}. \citet{goossens2009} investigated the forces that drive these waves and found that the magnetic tension force always dominates the pressure force. In addition \citet{goossens2009} showed that compression is small in the particular case of thin tubes. Hence these waves have not the typical properties of fast magneto-sonic waves and behave more as Alfv\'en waves. This lead \citet{goossens2009} to call these waves Alfv\'enic. The adjective Alfv\'enic was used in order to point out that pure Alfv\'en waves as decribed by \citet{alfven1942} can only exist in a uniform plasma of infinite extent.  \citet{goossens2011} reconsidered these waves in their section on quasi-modes and decided to call them surface Alfv\'en waves.

In the present paper we continue the theoretical investigation of the nature of the waves. We are concerned with vorticity since it is a quantity that is unequivocally related to Alfv\'en waves. Pure Alfv\'en waves in a homogeneous medium are the sole waves that represent vorticity perturbations and any spatial distribution of vorticity can be represented by means of Alfv\'en waves. Here we shall argue clearly that, in a piecewise homogeneous model, whether it is in the planar or cylindrical case, the  Alfv\'en surface waves have zero vorticity everywhere except at the discontinuity where all vorticity is concentrated. They are hence appropriately termed `Alfv\'en' surface waves and complement the pure Alfv\'en waves which are bound to the separate homogeneous regions. In view of their properties, the fundamental radial eigenmodes in a magnetic cylinder with $m\neq 0$ and phase velocities in the range $\vai$ to $\vae$ can be considered as surface Alfv\'{e}n waves. As a matter of fact, \citet{wentzel1979} was more ambitious and called all MHD waves  with phase velocities between $\vai$ and $\vae$ surface Alfv\'en waves. However, since the radial overtones have an oscillatory spatial behavior in the internal plasma we prefer to keep the term surface to the fundamental radial modes only. In addition we note that \citet{ionson1978}  called the surface waves Alfv\'enic probably to distinguish them from the pure Alfv\'en waves as described by \citet{alfven1942}. 

% In the present investigation we are concerned with vorticity. In particular we show that surface Alfv\'en waves for a true discontinuity in Alfv\'en velocity have zero vorticity in contrast to the classic Alfv\'en waves in a uniform plasma, except at the discontinuity where all vorticity is concentrated. We show that this result also applies to the radially fundamental eigenmodes in a magentic cylinder with $m\neq 0$ and phase velocities in the range $\vai$ to $\vae$ \citep[see Fig. 4 of][]{edwin1983}. These modes were called fast waves by \citet{edwin1983}. An important objective of the present paper is to show that these modes can be better considered as surface Alfv\'{e}n waves.

% and with the behavior of the part of the spectrum with phase velocities between the internal and the exteral Alfv\'en velocities in Figure 4 of \citet{edwin1983}. These modes were called fast waves by \citet{edwin1983}. An important objective of the present paper is to show that the radially fundamental eigenmodes with $m\neq 0$ and phase velocities in the range $\vai$ to $\vae$ can be considered as Alfv\'{e}n waves modified by non-uniformity.

%  On a true discontinuity, whether it is in the planar or cylindrical case,  We show that this result is not unique for a true discontinuity and also applies to the radially fundamental eigenmodes in a magnetic cylinder with $m\neq 0$ and phase velocities in the range $\vai$ to $\vae$. The main objective of the present paper is to show that, in view of their properties, these modes can be considered as surface Alfv\'{e}n(ic) waves.

The current paper is relevant to the recent discussion in the solar physics community on the nature of the observed ubiquitous, transverse waves \citep{tomczyk2007,depontieu2007,okamoto2007,mcintosh2011}. The original authors of these papers all claimed the detection of the Alfv\'en waves in the solar corona. This claim was challenged by \citet{vd2008} who argued that the waves should be interpreted as fast kink ($m=1$) waves following the nomenclature of \citet{edwin1983}. The results of \citet{goossens2009} and the present paper show that these waves are actually surface Alfv\'en waves.

% As mentioned in the previous paragraph, \citet{goossens2009} further investigated the nature of these waves. For the fundamental radial modes of the non-axisymmetric waves ($m\neq0$) in magnetic cylinders \citet{goossens2009} showed that compression is small in the particular case of thin tubes and that the magnetic tension force is the dominant restoring force. Hence these waves have not the typical properties of fast waves. Instead, they should be called Alfv\'enic since their properties are closer to those of the classic Alfv\'en waves.

% We are not the first ones to propose the name surface Alfv\'en waves to describe these waves. Actually \citet{wentzel1979} was more ambitious and called all MHD waves  with phase velocities between the internal and the external Alfv\'en velocities surface Alfv\'en waves. However, since the radial overtones have an oscillatory spatial behavior in the internal plasma we prefer to keep the term surface to the fundamental radial modes only.  \citet{ionson1978} also called the surface waves Alfv\'enic probably to distinguish them from the pure Alfv\'en waves as described by \citet{alfven1942}. As mentioned before, \citet{goossens2009} used the adjective Alfv\'enic and \citet{goossens2011} called these MHD waves surface Alfv\'en waves in their discussion on quasi-modes. Here we shall give additional arguments in favor of the name surface Alfv\'en waves.

% In particular, we focus on the computation and behaviour of the vorticity, since this is the quantity that describes Alfv\'en modes the best.

This paper is organized as follows. In Section~\ref{sec:uniform} we briefly review the properties of Alfv\'{e}n waves and slow and fast magneto-sonic waves in a uniform plasma of infinite extent with a constant magnetic field. Then we try to understand how the properties of the MHD waves known for uniform plasmas of infinite extent are modified when the plasma is no longer uniform and/or confined to a finite volume. First, in Section~\ref{sec:disc} this is done for the case of surface Alfv\'en waves in a true discontinuity in the Alfv\'en velocity. Later, in Section~\ref{sec:nonuniform} we extend our investigation to the case of MHD waves in non-uniform magnetic cylinders and show that the properties of the radially fundamental non-axisymmetric transverse waves in cylinders are remarkably similar to those of surface Alfv\'en waves in a true discontinuity. Finally, Section~\ref{sec:conclusions} contains our discussion and the relevant conclusions of this work. Those readers interested in our results but that do not wish to go in detail through the mathematical derivations will probably find useful the summary and discussion of Section~\ref{sec:conclusions}.

%%%%%%%%%%%%%%%%%%%%%%%%%%%%%%%%%%%
%%%%%%%%%%%%%%%%%%%%%%%%%%%%%%%%%%%
%%%%%%%%%%%%%%%%%%%%%%%%%%%%%%%%%%%

\section{MHD waves in a uniform plasma of infinite extent}
\label{sec:uniform}

The basic equations for the discussion of linear ideal MHD waves superimposed on a static plasma are
\begin{eqnarray}
\rho \frac{\partial {\bf v}}{\partial t} &=& - \nabla p' + \frac{1}{\mu} \left( \nabla \times {\bf B}' \right) \times {\bf B}+ \frac{1}{\mu} \left( \nabla \times {\bf B} \right) \times {\bf B}', \nonumber \\
\frac{\partial {\bf B}'}{\partial t} &=& \nabla \times \left( {\bf v} \times {\bf B} \right), \nonumber \\
\frac{\partial p'}{\partial t}+{\bf v}\cdot\nabla p &=& -\gamma p \nabla \cdot {\bf v}, \nonumber \\
 \frac{\partial \rho'}{\partial t}+{\bf v}\cdot\nabla\rho &=& - \rho \nabla \cdot {\bf v}, \label{basiceq}
\end{eqnarray}
where $p$, $\rho$, and ${\bf B}$ are the equilibrium plasma pressure, density, and magnetic field, respectively. We take $p$ and $\rho$ uniform and $\bf B$ straight and constant, namely ${\bf B} = B\, {\bf 1}_z$. In addition, $p'$, $\rho'$, ${\bf v}$, and ${\bf B}'$ are the Eulerian perturbations of the plasma pressure, density, velocity, and magnetic field, respectively, $\gamma$ is the adiabatic index, and $\mu$ is the magnetic permittivity. Equations~(\ref{basiceq}) assume that there are no equilibrium flows. For incompressible motions we have to replace the 3th equation of Equations~(\ref{basiceq}) with 
\begin{equation}
\nabla \cdot {\bf v} = 0, 
\end{equation}
and treat the plasma pressure perturbation, $p'$, as an unknown function in addition to the three unknown components of velocity.

%%%%%%%%%%%%%%%%%%%%%%%%%%%%%%%%%%%
%%%%%%%%%%%%%%%%%%%%%%%%%%%%%%%%%%%
%%%%%%%%%%%%%%%%%%%%%%%%%%%%%%%%%%%

In what follows we shall find it convenient to use the Lagrangian displacement, $\xii$, so that for a static plasma up to linear order
\begin{equation}
{\bf v} = \frac{\partial \xii}{\partial t},
\label{v-xi}
\end{equation}
and the Eulerian perturbation of total pressure, $P' = p' + p'_{\rm m}$, where $p'$ is the perturbation of the gas pressure and $p'_{\rm m} = {\bf B} \cdot {\bf B}' / \mu$ is the perturbation of the magnetic pressure.

{\bf With the use of the Lagrangian displacement, $\xii$, one may integrate the induction equation, the energy equation and the continuity equation at once and use these expressions to eliminate all other variables from the momentum equation so that it takes the form:}
\begin{equation}
 \rho \frac{\partial ^2\xii}{\partial t^2}  = \mathbf{F}(\xii)\ .
\end{equation}
{\bf Here $\mathbf{F}$ is the force operator as derived first by \citet{bernstein1958}, \citep[see also][]{frieman1960,goedbloed1983}.}

We consider a uniform plasma of infinite extent. We study linear planar harmonic  waves and put the wave variables proportional to the factor 
\begin{equation}
\exp(i {\bf k}\cdot {\bf r}- i \omega t)
\label{PlaneWaves}
\end{equation}
where ${\bf k} = \left(k_x, k_y,k_z\right)^t $ is the wave vector, ${\bf r}=\left(x,y,z  \right)$ is the position vector, and $\omega$ the frequency. Variables that catch the basic physics of the waves \citep[see, e.g.,][]{thompson1962, goossens2003}  are the component of the displacement parallel to the equilibrium magnetic field, $\xi_z$, the compression, $Y$, and the component of vorticity along the magnetic field, $Z$. $Y$ and $Z$ are defined as 
\begin{eqnarray}
Y &=& {\bf k} \cdot \xii = - i \nabla \cdot \xii, \nonumber \\
 Z &=& \left( {\bf k} \times \xii \right) \cdot \mathbf{1}_z = - i \left( \nabla \times \xii \right)_z.
\label{YZ}
\end{eqnarray}
With the use of these variables the equations that govern the waves are
\begin{eqnarray}
\omega^2 \xi_z - k_z \vs^2  Y  & = & 0, \nonumber \\
k^2 \va^2 k_z  \xi_z +  \left( \omega^2 - k^2 (\vs^2 + \va^2) \right) Y & = & 0, \nonumber \\
\left( \omega^2 - \omegaA^2 \right) Z & = & 0 . \label{FSAW}
\end{eqnarray}
Here $k^2 = k_x^2 + k_y^2 + k_z^2$, $\vs^2 = \gamma p/ \rho$ is the square of the speed of sound, $\va^2 = B^2 / \mu \rho$ is the square of the Alfv\'en velocity, and $\omegaA = k_z \va$ is the Alfv\'en frequency. The system of Equations~(\ref{FSAW}) is decoupled into two subsystems. The first two equations  involve the variables $\xi_z$ and $Y$ but not $Z$. They are the equations that define the (slow and fast) magneto-sonic waves. These waves have compression and a component of velocity parallel to the magnetic field, but no vorticity.  The third equation only contains $Z$. It defines the Alfv\'{e}n waves. Alfv\'{e}n waves  have vorticity but  no compression and no component of velocity parallel to the magnetic field.

\subsection{Alfv\'{e}n waves} 
The dispersion relation of Alfv\'en waves is obtained from Equations~(\ref{FSAW}) with $\xi_z = 0$, $Y=0$, and $Z\neq0$. Thus the dispersion relation of Alfv\'en waves is 
\begin{equation}
\omega = \omegaA =  k_z \va.
\label{DispersionAW}
\end{equation}
An Alfv\'{e}n wave exists for any wave vector ${\bf k}$ but its frequency only depends on the component of  ${\bf k}$ parallel to the equilibrium magnetic field.  The frequency is degenerate with respect to the components $k_x$ and $k_y$  of the wave vectors in planes normal to the equilibrium magnetic field lines. Since frequency only depends on $k_z$, the group velocity, ${\bf v}_{\rm gr}$, is always directed along the equilibrium magnetic field and is equal to $\va$, while the phase velocity, ${\bf v}_{\rm ph}$, is by definition directed along the wave vector ${\bf k}$, namely ${\bf v}_{\rm ph} = \va \cos(\theta) {\bf 1_k}$, where $\theta$ is the angle that  ${\bf k}$ makes with ${\bf B}$, and ${\bf 1_k}$ and ${\bf 1_B}$ are the unit vectors in the direction of ${\bf k}$ and ${\bf B}$, respectively. 
Alfv\'{e}n waves are anisotropic in the extreme. We recall that in our discussion so far no particular choice has been made for ${\bf k}$ and the wave vector is general.

%  For the Alfv\'{e}n waves under study and with the prescription of Equation~(\ref{PlaneWaves}), ${\bf v}$ and $\xii$ are related as 
% ${\bf v} = - i \omegaA \xii_{\rm A}$ and  $\xii_{\rm A}$ is given by
% \begin{eqnarray}
% \xii_{\rm A} & = & \xii_{\rm A,0}\exp \left[ i \left(k_x x + k_y y + k_z z \right) - i \omegaA t \right], \nonumber \\
% \xii_{\rm A,0} & = &  \left(-k_y,  k_x, 0 \right)^t  \frac{\xi_{0,y}}{\sqrt{k_x^2 + k_y^2}}.
% \label{STV-AW}
% \end{eqnarray}
% $\xii_{\rm A,0} $ is a constant horizontal vector perpendicular to the horizontal  wave vector ${\bf k}_{\rm h} = (k_x, k_y,0)^t$ so as to make compression zero and vorticity maximal. 
% As $P' = 0$ for Alfv\'{e}n waves, the only restoring force is the magnetic tension force,
% \begin{equation}
% {\bf T}_{\rm A} = - \rho  \omegaA^2 \xii_{\rm A}.
% \label{TensionFAW}
% \end{equation}
% In our discussion so far no particular choice has been made for ${\bf k}$ and the wave vector is general. 

\subsection{Magneto-sonic waves} 
The dispersion relation for the slow and fast magneto-sonic waves is readily obtained by imposing that the subsystem in Equations~(\ref{FSAW}) for the variables $\xi_z$ and $Y$ has a non-trivial solution.   The result is
\begin{eqnarray}
\omega^2_{\rm f,sl} & = & \frac{k^2 (\vs^2 + \va^2)}{2} \left[ 1 \pm \left(1 - \frac{4 \omegaC^2}{k^2 (\vs^2 + \va^2)} \right)^{1/2} \right], \nonumber \\
\omegaC^2  & = &  \frac{\vs^2}{\vs^2 + \va^2} \omegaA^2
\label{DispersionFSW}
\end{eqnarray}
where the $\pm$ sign corresponds to the fast/slow magneto-sonic waves, respectively, and $\omegaC$ is the cusp frequency. Subscripts ``f'' and ``sl'' denote quantities related to fast and slow waves, respectively. Let $\beta$ denote the ratio of the plasma pressure to the magnetic pressure. Then in the approximation $\beta \to 0$ ($\vs=0$), the slow waves disappear from the scene, i.e., $\omega_{\rm sl} = 0$. Fast waves still remain in the zero-$\beta$ approximation with frequency $\omega_{\rm f} = k \va$. On the contrary, for $\vs \to \infty$ fast waves are removed to infinite frequencies and a particular form of slow waves remain with $\omega_{\rm sl} = \omegaA$.

\subsection{Forces and motions}
It is instructive to  write the equation of motion as 
\begin{eqnarray}
\rho \frac{\partial ^2 \xii_{\rm h}}{\partial t ^2} &=& - \nabla_{\rm h} P' + {\bf T}_{\rm h}, \nonumber \\
\rho \frac{\partial ^2 \xi_z}{\partial t ^2} &=& - \frac{\partial p'}{\partial z},
\label{Eq-Motion1}
\end{eqnarray}
where $\bf T_{\rm h}$ is (the horizontal component of) the magnetic tension force and $\nabla_{\rm h} $ and  $\xii_{\rm h} = (\xi_x, \xi_y, 0)^t $ are the gradient operator and the displacement, respectively, in horizontal planes perpendicular to the constant vertical magnetic field. The component of the displacement parallel to the magnetic field, $\xi_{z}$, is solely driven by plasma pressure and unaffected by magnetic forces. With the temporal and spatial dependence specified by Equation~(\ref{PlaneWaves}), and ${\bf k_h}= (k_x, k_y, 0)^t$  Equations~(\ref{Eq-Motion1}) can be written as
\begin{eqnarray}
- \rho  \omega^2  \xii_{\rm h} & = & - i {\bf k}_{\rm h} P' + {\bf T}_{\rm h} \nonumber \\
{\bf T}_{\rm h} & = & - \rho \omegaA^2 \xii_{\rm h}, \nonumber \\
\rho \omega^2  \xi_{z} & = & i k_{z}  p'.
\label{Eq-Motion2}
\end{eqnarray}
The force in horizontal planes due to total pressure, ${\bf \Pi} $, is
\begin{equation}
{\bf \Pi} = - i {\bf k}_{\rm h} P' = -\rho (\omega^2 - \omegaA^2) \xii_{\rm h}.
\label{P-force}
\end{equation}
Hence the ratio of the horizontal components of total pressure force to the corresponding 
components of the magnetic tension force is
\begin{equation}
\Lambda(\omega^2) =\frac{ \omega^2 - \omegaA^2}{  \omegaA^2 }=\frac{\omega^2}{\omegaA^2}-1.
\label{RatioP-T2}
\end{equation}
Equation~(\ref{RatioP-T2}) is general and applies to Alfv\'{e}n waves with $\omega^2 = \omega_A^2$, 
fast waves with  $\omega^2 = \omegaf^2$ and to slow waves with 
$\omega^2 = \omegasl^2$.

For the Alfv\'{e}n waves $\Lambda(\omegaA^2) = 0 $ so that the only restoring force is the magnetic tension force.  The displacements are incompressible and confined to horizontal planes since $\xi_z = 0$. For the Alfv\'en waves the displacement $\xii =  \xii_{\rm h}$ and is perpendicular to the horizontal wave vector ${\bf k}_{\rm h}$ so as to make compression zero and vorticity maximal. Alfv\'{e}n waves are highly anisotropic and totally insensitive to the value of the sound speed of the plasma.

On the contrary, magneto-sonic waves involve both the total pressure force and the magnetic tension force. It is now appropriate to consider the relative importance of horizontal compression ${\bf k}_{\rm h}\cdot\xii_{\rm h}$ and longitudinal compression $k_z\xi_z$, which can be readily derived from the first line of Equation~(\ref{FSAW}): 
\begin{equation}
\frac{{\bf k}_{\rm h}\cdot\xii_{\rm h}}{k_z\xi_z} = \frac{\omega^2 - k_z^2 \vs^2}{k_z^2\vs^2} =\frac{\omega^2}{k_z^2\vs^2} - 1.
\label{Ratio-compP-compL}
\end{equation}
Equation~(\ref{Ratio-compP-compL}) is again applicable to fast waves with $\omega^2 = \omegaf^2$ and to slow waves with $\omega^2 = \omegasl^2$. For Alfv\'{e}n waves with $\omega^2 = \omega_A^2$, it is largely irrelevant as both the denominator and the numerator vanish and no compression is involved at all. The equation tells us that the coupling of the longitudinal compression to the perpendicular compression depends on the sound speed. For fast waves the ratio is positive while it is negative for slow waves. This means that compression, $Y$ (the sum of the denominator and numerator of (\ref{Ratio-compP-compL})), is maximized for fast waves and reduced for slow waves. If the sound speed is small, the longitudinal dynamics decouples and is irrelevant to the perpendicular dynamics. The slow waves vanish and there are no longitudinal motions in the fast waves. On the contrary if the sound speed is large, the above ratio approaches $-1$, illustrating the perfect coupling in an incompressible medium. The fast waves are banned to infinity while for the slow modes $\omegasl^2\rightarrow\omegaA^2$ so that the dynamics in the perpendicular direction is dominated by the magnetic tension force.

%  Fast waves are driven by both the total pressure force and the magnetic tension force. The displacements are compressible and have no vorticity. In horizontal planes they are parallel to the horizontal wave vector and the vertical component of the displacement combines with the horizontal components so as to make compression maximal.  Fast waves are sensitive to the value of the sound speed. 
% 

%%%%%%%%%%%%%%%%%%%%%%%%%%%%%%%%%%%
%%%%%%%%%%%%%%%%%%%%%%%%%%%%%%%%%%%
%%%%%%%%%%%%%%%%%%%%%%%%%%%%%%%%%%%

\section{MHD waves on a true discontinuity}
\label{sec:disc}

In the previous section we have studied the properties of Alfv\'{e}n waves and magneto-sonic waves in a uniform plasma of infinite extent. In that configuration, Alfv\'{e}n waves are strikingly different from fast waves. 

Let us now see how a deviation from the uniform plasma of infinite extent adds new complexity. 
First of all consider a magnetic field that is still straight and unidirectional throughout space and let all equilibrium variation depend only on one cartesian coordinate $x$ which is directed perpendicular to the equilibrium magnetic field. Now the wave variables can be put proportional to $\exp \left( ik_y y + ik_z z - i \omega t \right)$, but the functional dependence on $x$ must remain unspecified. The relevant equations become:
\begin{eqnarray}
 \rho \left( \omega^2 - \omegaA^2  \right)\frac{{\rm d} \xi_x}{{\rm d}x} &=& K^2  P', \nonumber \\
 \rho \left( \omega^2 - \omegaA^2  \right) \xi_x  &=& \frac{{\rm d}P'}{{\rm d}x}, \nonumber \\
 \rho \left( \omega^2 - \omegaA^2  \right) \xi_y  &=& i k_y P', \nonumber \\
 \rho (\vs^2 + \va^2)\left( \omega^2 - \omegaC^2  \right) \xi_z  &=& i k_z \vs^2P', \label{eq:compequations} 
\end{eqnarray}
where
\begin{equation}
  K^2 =- \frac{\omega^4 -(\vs^2 + \va^2)(\omega^2 -\omegaC^2)(k_y^2 + k_z^2)}{(\vs^2 + \va^2)\left( \omega^2 - \omegaC^2  \right)},
\end{equation}
is a function of $x$. 

% In order to focus on Alfv\'{e}n waves we consider incompressible motions, i.e., $\nabla \cdot \xii = 0$.  The relevant equations for incompressible motions on a one-dimensional (1D) plasma equilibrium are
% \begin{eqnarray}
%  \rho \left( \omega^2 - \omegaA^2  \right) \frac{{\rm d} \xi_x}{{\rm d}x} &=& k^2  P', \nonumber \\
%  \rho \left( \omega^2 - \omegaA^2  \right) \xi_x  &=& \frac{{\rm d}P'}{{\rm d}x}, \nonumber \\
%  \rho \left( \omega^2 - \omegaA^2  \right) \xi_y  &=& i k_y P', \nonumber \\
%  \rho \left( \omega^2 - \omegaA^2  \right) \xi_z  &=& i k_z P',\nonumber \\
%  \rho \omega^2 \xi_z &=& i k_z p', \label{eq:incequations} 
% \end{eqnarray}
% where all quantities have the same meaning as before.
% \begin{equation}
%  \rho \left( \omega^2 - \omegaA^2  \right) \frac{{\rm d}}{{\rm d}x} \left( \frac{1}{\rho \left( \omega^2 - \omegaA^2  \right) } \frac{{\rm d} P'}{{\rm d}x} \right) = k^2 P'. \label{eq:pinco}
% \end{equation}

We may combine the first two equations in Equations~(\ref{eq:compequations}) into a second order ordinary differential equation for $P'$, namely
\begin{equation}
 \rho \left( \omega^2 - \omegaA^2  \right) \frac{{\rm d}}{{\rm d}x} \left( \frac{1}{\rho \left( \omega^2 - \omegaA^2  \right) } \frac{{\rm d} P'}{{\rm d}x} \right) = K^2 P'. \label{eq:pcomp}
\end{equation}
From Equations~(\ref{eq:compequations}) we can also compute the components of vorticity, $\nabla \times \xii$, in the $z$-direction as
\begin{equation}
 \left( \nabla \times  \xii \right) \cdot \mathbf{1}_z = i k_y P' \frac{{\rm d}}{{\rm d}x} \left( \frac{1}{\rho \left( \omega^2 - \omegaA^2 \right)}\right). \label{eq:vortint}
\end{equation}
We shall use these equations in the following subsections.

Here we will first consider an equilibrium consisting of two uniform plasmas separated by a sharp discontinuity in the Alfv\'en velocity. The discontinuity coincides with the plane $x = x_0$. For simplicity, we take a constant magnetic field along the $z$-direction. Thus, the discontinuity in the Alfv\'en velocity is introduced by a discontinuity in density as,
\begin{equation}
\rho(x) = \left\{ \begin{array}{lll}
        \rhoi, & \textrm{if}, & x \leq x_0, \\
 \rhoe, & \textrm{if}, & x > x_0,
       \end{array} \right.
\end{equation}
where both $\rhoi$ and $\rhoe$ are constants and $\rhoi \neq \rhoe$. The MHD waves of this plasma configuration have been studied before \citep[see, e.g.,][]{Wentze1979planar,roberts1981}. The aim here is to point out the differences between classic Alfv\'en waves and surface Alfv\'en waves.

The basic Equations~(\ref{FSAW}) are valid at both sides of the interface and $K^2$ (whenever it is negative) in Equation~(\ref{eq:pcomp}) must be indentified with $-k^2_x$ used earlier. The only difference is hence in the boundary conditions. In particular, the solutions at both sides of the discontinuity need to be matched together by ensuring continuity of both total pressure $P'$ and normal displacement $\xi_x$. The interaction between the two media allows for solutions in that range where $K^2>0$. In that case solutions are given by:
\begin{equation}
 P' (x) = A_\mathrm{i,e} \exp \left( \pm K_\mathrm{i,e} \left( x - x_0 \right) \right).
\end{equation}
So whenever $K^2>0$ the modes correspond to surface waves decaying away from the surface and their existence crucially involves the interaction between the two media. In contrast, the classical slow and fast waves discussed earlier correspond to solutions in the domain where $K^2<0$.

% \begin{eqnarray}
%   \rho \left( \omega^2 - \omegaA^2  \right)  \left( \nabla \times  \xii \right) &\cdot& \mathbf{1}_z = \nonumber \\
%  &-&  \frac{i k_y P'}{\rho \left( \omega^2 - \omegaA^2  \right) } \frac{{\rm d}}{{\rm d}x} \left(  \rho \left( \omega^2 - \omegaA^2  \right) \right), \\
%   \rho \left( \omega^2 - \omegaA^2  \right)  \left( \nabla \times  \xii \right) &\cdot& \mathbf{1}_y = \nonumber \\
%  &&  \frac{i k_z P'}{\rho \left( \omega^2 - \omegaA^2  \right) } \frac{{\rm d}}{{\rm d}x} \left(  \rho \left( \omega^2 - \omegaA^2  \right) \right).
% \end{eqnarray}

\subsection{Classic Alfv\'en waves}
\label{sec:discclassic}
However, let us first retrieve the classic Alfv\'en waves by requiring $P'=0$ and $\xi_z = 0$ everywhere. Hence also $p'=0$ and since $\nabla \cdot \xii =0$ also $\rho'=0$. We find two different solutions under these conditions. The first solution has frequencies $\omega^2 = \omegaAi^2$ and motions satisfying
\begin{equation}
\begin{array}{lll}
        \left( \nabla \times  \xii \right) \cdot \mathbf{1}_z  \neq 0, & \textrm{for}, & x \leq x_0, \\
  \left( \nabla \times  \xii \right) \cdot \mathbf{1}_z = 0, & \textrm{for}, & x > x_0.
       \end{array} 
\end{equation}
The second solution has frequencies $\omega^2 = \omegaAe^2$ and motions satisfying
\begin{equation}
\begin{array}{lll}
        \left( \nabla \times  \xii \right) \cdot \mathbf{1}_z  = 0, & \textrm{for}, & x \leq x_0, \\
  \left( \nabla \times  \xii \right) \cdot \mathbf{1}_z \neq 0, & \textrm{for}, & x > x_0.
       \end{array} 
\end{equation}
Hence, classic Alfv\'en waves with different  Alfv\'en frequencies $\omegaAi^2$ and $\omegaAe^2$ are confined to the half spaces $x \leq x_0$ and $x > x_0$, respectively.

% The first solution has frequencies $\omega^2 = \omegaAi^2$ and motions satisfying
% \begin{equation}
% \begin{array}{llll}
%         \xi_z \neq 0, & \left( \nabla \times  \xii \right) \cdot \mathbf{1}_y  \neq 0, & \textrm{for}, & x \leq x_0, \\
%  \xi_z = 0, & \left( \nabla \times  \xii \right) \cdot \mathbf{1}_y = 0, & \textrm{for}, & x > x_0.
%        \end{array} 
% \end{equation}
% The second solution has frequencies $\omega^2 = \omegaAe^2$ and motions satisfying
% \begin{equation}
% \begin{array}{llll}
%        \xi_z = 0, & \left( \nabla \times  \xii \right) \cdot \mathbf{1}_y  = 0, & \textrm{for}, & x \leq x_0, \\
%   \xi_z \neq 0, & \left( \nabla \times  \xii \right) \cdot \mathbf{1}_y \neq 0, & \textrm{for}, & x > x_0.
%        \end{array} 
% \end{equation}
% Hence, slow waves with different  slow frequencies (equal to the Alfv\'en waves) $\omegaAi^2$ and $\omegaAe^2$ are confined to the half spaces $x \leq x_0$ and $x > x_0$, respectively.

\subsection{Surface Alfv\'en waves}
\label{sec:discsurf}
Let us now turn to the motions that do involve $P' \neq 0$. We have argued above that the classical slow and fast waves are related to the domain where $K^2<0$ while if 
$K^2>0$ the modes are surface waves relying on the interaction between the two media. The roots of $K^2$ as a function of frequency are easily determined as they correspond to the fast and slow frequencies (\ref{DispersionFSW}) with $k_x=0$. Additionally $K^2$ changes sign at the cusp frequency. This allows to easily indentify the frequency ranges for the (classical) slow and fast modes and of the surface modes. Now consider two important limiting cases. In the incompressible limit the slow frequencies collapse to the Alfv\'en frequency and the fast frequencies are banned to infinity, hence only surface type solutions remain and in fact $K^2\approx (k^2_z+k^2_y)$. Similarly, in what is called the limit of `nearly perpendicular propagation' ($k^2_y\gg k^2_z$) where $K^2\approx k^2_y$. {\bf \citet{Wentze1979planar} already remarked that the incompressibility condition is a good approximation when $k^2_y\gg k^2_z$. Furthermore,} the latter approximation is highly relevant to thin cylindrical tubes{ \bf considered further in this paper}. In both cases a solution to the dispersion relation expressing the matching of $\xi_x$ and $P'$ at $x = x_0$ is found as:
\begin{equation}
 \omega^2 = \frac{\rhoi\omegaAi^2+\rhoe\omegaAe^2 }{\rhoi+\rhoe} \equiv \omega_{\rm k}^2. \label{eq:kinkfreq}
\end{equation}
The frequency lies in between the internal and external Alfv\'en frequency and hence by Equation~(\ref{RatioP-T2}) $\Lambda$ can't be very large at neither side of the discontinuity. In fact, if the magnetic field is constant, $-1<\Lambda<1$ so that the perpendicular dynamics is always dominated by the tension force.
\\
One might be tempted to classify the wave as either fast or slow, since there is compression involved. In that case one would need to take the ratio defined in Equation~(\ref{Ratio-compP-compL}) into account. In the incompressible limit we would then necessarily be dealing with a slow mode while in a cold plasma it would need to be interpreted as a fast mode. However, in the limit of nearly perpendicular propagation the perpendicular dynamics is completely insensitive to the sound speed. The mode is dominated by the perpendicular dynamics and the sound speed just influences to what extent this couples to longitudinal motions as well. The insensitivity to the sound speed and the dominance of tension as a driving force is a strong argument to call these modes Alfv\'enic.

%  as it is seen in their expressions,
% \begin{eqnarray}
%  \xi_y (x) &=& \pm \frac{i k_y P'}{\rho \left( \omega^2 - \omegaA^2 \right)}, \nonumber \\
% \xi_z (x) &=& \pm \frac{i k_z P'}{\rho \left( \omega^2 - \omegaA^2 \right)},
% \end{eqnarray}
% where $+$ sign is for $x \leq x_0$ and the $-$ sign is for $x > x_0$, and the local values of $\rho$ and $\omegaA^2$ must be used as appropriate.

Let us now focus on vorticity. From Equation~(\ref{eq:vortint}) we can evaluate the $z$-component of vorticity of the surface Alfv\'en wave. To do so we note that $\rho \left( \omega^2 - \omegaA^2 \right)$ is a piecewise constant in our equilibrium and can be expressed as
\begin{equation}
 \frac{1}{\rho \left( \omega^2 - \omegaA^2 \right)} = \frac{\rhoi + \rhoe}{\rhoi \rhoe}\frac{1}{\omegaAe^2 - \omegaAi^2 } \left( 1 - 2 H(x-x_0) \right),
\end{equation}
where $H(x- x_0)$ is the Heaviside function defined as
\begin{equation}
H(x - x_0) = \left\{ 
 \begin{array}{lll}
    1, & \textrm{if}, & x > x_0, \\
  0, & \textrm{if}, & x < x_0.
       \end{array} \right. \label{eq:heviside}
\end{equation}
Thus it is straightforward to write the expression for $ \left( \nabla \times  \xii \right) \cdot \mathbf{1}_z$ as
\begin{equation}
 \left( \nabla \times  \xii \right) \cdot \mathbf{1}_z = - 2 i k_y  P' \frac{\rhoi + \rhoe}{\rhoi \rhoe} \frac{1}{\omegaAe^2 - \omegaAi^2 } \delta(x-x_0), \label{eq:vortdelta}
\end{equation}
where $\delta(x-x_0)$ is the delta function defined as
\begin{equation}
\delta(x - x_0) = \left\{ 
 \begin{array}{lll}
    1, & \textrm{if}, & x = x_0, \\
  0, & \textrm{if}, & x \neq x_0.
       \end{array} \right. \label{eq:delta}
\end{equation}
Equation~(\ref{eq:vortdelta}) shows that the surface Alfv\'en wave is different from the classic Alfv\'en wave. For the classic Alfv\'en wave $ \left( \nabla \times  \xii \right) \cdot \mathbf{1}_z$ is different from zero everywhere in the appropriate half space $x<x_0$ or $x>x_0$. The surface Alfv\'en wave has  $ \left( \nabla \times  \xii \right) \cdot \mathbf{1}_z = 0$ everywhere except at the discontinuity $x = x_0$.

When the true discontinuity in $\va$ is replaced with a continuous variation from $\vai$ to $\vae$ then the interval $[\omegaAi, \omegaAe]$ is filled with the continuous spectrum of shear Alfv\'en waves. Each magnetic surface, i.e., a surface of constant $\va$, now oscillates at its local Alfv\'en frequency. The frequency of the surface Alfv\'en wave is in the Alfv\'en continuous spectrum and gets damped by resonant absorption. This happens primarily at the resonant position, $x_{\rm A}$, where the local Alfv\'en frequency equals the frequency of the surface wave. \citep[see, e.g.,][]{tataronis1973,grossmann1973,ionson1978,hasegawauberoi,goedbloed1983,hollweg1987a,hollweg1987b,hollweg1988,rudermangoossens1993, rudermangoossens1996,rudermanTG1995}. Vorticity is no longer confined to a single surface but is spread out over the whole region of non-uniformity as indicated by Equation~(\ref{eq:vortint}). 

\section{MHD waves in a magnetic cylinder}
\label{sec:nonuniform}
In the present section we are interested in MHD waves superimposed on a 1D straight cylinder. First we consider a piece-wise constant density as in \citet{edwin1983}. Later we replace the jump in density with  a continuous variation of density. The properties of kink MHD waves in 1D straight cylinders were discussed by \citet{goossens2009}. As for the case of a true discontinuity, here we give special attention to vorticity.  We shall show that the surface Alfv\'en waves in a true discontinuity and the radially fundamental MHD waves with phase velocities between $\vai$ and $\vae$ in a cylinder have strikingly similar properties.

\subsection{General theory}

The equilibrium configuration is a straight cylinder. We use cylindrical coordinates, namely $r$, $\varphi$, and $z$, for the radial, azimuthal, and longitudinal coordinates, respectively. The equilibrium quantities are functions of $r$ only. Since the background model is independent of the spatial coordinates $\varphi$ and $z$, and of time, $t$, the perturbed quantities are put proportional to 
\begin{equation}
\exp \left(i (m \varphi + k_z z - \omega t) \right),
\label{mkt}
\end{equation}
where $m$ is the azimuthal wave number, $k_z$ is the longitudinal wave number, and $\omega$ is the frequency as before. Since the background is variable in the radial direction there is not really a radial wave number. Alternatively, we can use the number of nodes in the radial part of the eigenfunctions to distinguish between radial fundamental and overtone modes. The equations for linear MHD motions superimposed on a 1D cylindrical equilibrium model can be found in, e.g., \citet{appert1974,sakurai1991,goossens1992,goossens1995}.    For a straight and constant magnetic field the equations for linear displacements on a 1D cylindrical equilibrium model take the following simplified form
\begin{eqnarray}
D\frac{{\rm d} (r \xi_r)}{{\rm d}r}&  = &  - C_2 r P', \nonumber \\
\frac{{\rm d} P'}{{\rm d}r} & = & \rho (\omega^2 - \omegaA^2) \xi_r, \nonumber \\
\rho (\omega^2 - \omegaA^2) \xi_{\varphi} & = & \frac{i m}{r} P', \nonumber \\
\rho (\omega^2 - \omegaC^2) \xi_{z} & = & i k_{z} \frac{\vs^2}{\vs^2 + \va^2} P', 
\label{MHDWavesODE1}
\end{eqnarray}
where all quantities have the same meaning as in previous sections. We recall that now the equilibrium quantities are functions of $r$. The coefficient functions $D$ and $C_2$ are
\begin{eqnarray}
\mbox{}\nonumber \\
D & = & \rho (\vs^2 + \va^2) (\omega^2 -\omegaA^2)(\omega^2- \omegaC^2), \nonumber \\
C_2 & = & \omega^4 -(\vs^2 + \va^2)(\omega^2 -\omegaC^2)(\frac{m^2}{r^2} + k_z^2).
\label{MHDwavesCF}
\end{eqnarray}

Since \citet{appert1974} it is well known that the system formed by Equations~(\ref{MHDWavesODE1}) has mobile regular singularities at the positions $r=\ra$ and $r=\rc$ where $\omega = \omegaA(\ra)$ and $\omega = \omegaC(\rc)$. This leads to the definition of Alfv\'{e}n and slow (or cusp)  continuous parts, namely 
\begin{equation}
[\min\omegaA, \max\omegaA], \qquad [\min\omegaC, \max\omegaC],
\label{Alfven-slow-continuum}
\end{equation}
with singular eigensolutions \citep[see, e.g.,][]{goedbloed1983,sakurai1991,goossens1995,tirry1996}. 
For a straight field the $\varphi$-  and  $z$-directions are the directions of constant Alfv\'en velocity perpendicular and parallel to the magnetic field lines, respectively. The $r$-direction
is normal to the surfaces of constant Alfv\'en velocity. Hence for a straight field $\xi_{\varphi} = \xi_{\perp}$ is the characteristic quantity for the Alfv\'{e}n waves and $\xi_z = \xi_{\parallel}$ for the slow waves, where $\parallel$ and $\perp$ denote the directions parallel and perpendicular to the equilibrium magnetic field, respectively. $\xi_r$ characterizes the fast magneto-sonic waves. 

For an equilibrium with a straight magnetic field, the Eulerian perturbation of total pressure $P'$ is the quantity that produces waves with mixed Alfv\'en and magneto-sonic properties. The coupling function, $C_{\rm A}$, is 
\begin{equation}
C_{\rm A} = \frac{m B}{  r} P'.
\label{CAStraightF}
\end{equation}
The coupling function $C_{\rm A}$ was introduced by \citet{sakurai1991} and \citet{goossens1992}, and its role was discussed by, e.g., \citet{goossens2008iau} and \citet{goossens2011}.

% See details of the importance of $C_{\rm A}$ in, e.g., \citet{sakurai1991, goossens1992}. For
% $m=0$ the coupling function is $C_{\rm A} = 0$. Hence the third equation in
% Equations~(\ref{MHDWavesODE1}) becomes decoupled from the remaining
% equations for $m=0$, i.e.,
% \begin{equation}
% \rho (\omega^2 - \omegaA^2) \xi_{\varphi}  =  0.
% \label{TorsionalAW}
% \end{equation}
% In order to have non-trivial solutions with $\xi_{\varphi} \neq 0$ it is required that $\omega^2 = \omegaA^2$. Since $\omegaA^2(r)$ is a function of position there is a continuous spectrum of Alfv\'{e}n waves.  This means  that we have pure Alfv\'{e}n waves for $m=0$ in a non-uniform cylindrical plasma with a straight field. The
% axi-symmetric MHD waves are decoupled in torsional Alfv\'{e}n
% waves and sausage magneto-sonic waves. For an axi-symmetric
% non-uniform 1D cylindrical plasma this is the only case
% where pure incompressible Alfv\'{e}n waves show up in the analysis. Each surface of constant Alfv\'en velocity oscillates with its own local Alfv\'{e}n frequency.
% The torsional Alfv\'{e}n waves have frequencies depending on position since $\omegaA(r)$ depends on position. 

The two first order differential equations in Equations~(\ref{MHDWavesODE1}) can be rewritten as a second order ordinary differential equation for $P'$ as
\begin{equation}
\frac{\rho (\omega^2 - \omegaA^2)}{r} \frac{\rm d }{{\rm d}r}\left[\frac{r}{\rho (\omega^2 - \omegaA^2)} \frac{{\rm d} P'}{{\rm d}r} \right] =
\left[ \frac{m^2}{r^2} - \Gamma(\omega^2) \right] P',
\label{MHDWavesODE2}
\end{equation}
where $\Gamma(\omega^2)$ is  defined as
\begin{equation}
\Gamma(\omega^2) = \frac{(\omega^2 - k_z^2 \vs^2)(\omega^2 - \omegaA^2)}{(\vs^2 + \va^2)(\omega^2 - \omegaC^2)}.
\label{GammaMHDR}
\end{equation}
Note that Equation~(\ref{Eq-Motion2}) is applicable here also with $\xii_{\rm h} = \xi_r \mathbf{1}_r + \xi_{\varphi} \mathbf{1}_{\varphi}$ and ${\bf k}_{\rm h} =  -\mathbf{1}_r  i\frac{{\rm d}}{{\rm d}r}+ \frac{m}{r} \mathbf{1}_{\varphi}$. Equations~(\ref{P-force}) and (\ref{RatioP-T2}) remain valid so that we have a simple expression for $\Lambda$ to decide on the relative contributions of pressure and tension forces. 

% \subsubsection{Vorticity}
% \label{sec:vort}

Recall that the component of vorticity parallel to the equilibrium magnetic field, $ (\nabla  \times \xii)\cdot \mathbf{1}_z$, is unequivocally related to Alfv\'{e}n waves. Alfv\'{e}n waves have $ (\nabla  \times \xii)\cdot \mathbf{1}_z \neq 0$ everywhere in an infinite and uniform plasma, while surface Alfv\'{e}n waves in a true discontinuity have $(\nabla  \times \xii)\cdot \mathbf{1}_z \neq 0$ at the discontinuity only. Hence for comparison with these previous cases and for later use it is instructive to derive an
equation for this quantity:  
\begin{eqnarray}
(\nabla \times \xii) \cdot \mathbf{1}_z  &=& i \frac{m}{r} P' \frac{\rm d}{{\rm d}r} \left(\frac{1}{\rho (\omega^2 - \omegaA^2)} \right). 
\label{Vorticity2}
\end{eqnarray}
Note that this equation is formally equivalent to Equation~(\ref{eq:vortint}) derived for waves on a true discontinuity if in Equation~(\ref{eq:vortint}) $k_y$ is replaced by $m/r$ and the derivative in $x$ is replaced by a derivative in $r$.

\subsection{Piecewise uniform plasma }
Here we consider the case studied by \citet{edwin1983}, i.e., the case of  a piecewise constant density profile. The general situation in which the density varies continuously is considered later.  We assume the following density profile,
\begin{equation}
\rho(r) = \left\{ \begin{array}{lll}
        \rhoi, & \textrm{if}, & r \leq R, \\
 \rhoe, & \textrm{if}, & r > R,
       \end{array} \right.
\end{equation}
where both $\rhoi$ and $\rhoe$ are constants and $R$ denotes the radius of the cylinder. Subscripts i and e refer to internal and external plasmas, respectively. We focus on the case $\rhoi \geq \rhoe$. Thus, there is a jump in density, namely $\rhoi-\rhoe$, at the cylinder boundary.

We rewrite Equations~(\ref{MHDWavesODE2}) and (\ref{Vorticity2}) when the density $\rho$ and the local Alfv\'{e}n frequency $\omegaA$ are both constants, namely 
\begin{eqnarray}
 \frac{1}{r}\frac{\rm d }{{\rm d}r} \left(r \frac{{\rm d} P'}{{\rm d}r} \right) &=& \left[ \frac{m^2}{r^2} - \Gamma(\omega^2) \right] P' , \label{MHDWavesODE3} \\
 \rho (\omega^2 - \omegaA^2)  (\nabla \times \xii) \cdot \mathbf{1}_z &=& 0
\label{MHDWavesODE4}
\end{eqnarray}
Equation~(\ref{MHDWavesODE3}) is Equation~(5) of \citet{edwin1983}. Equation~(\ref{MHDWavesODE4}) is Equation~(3b) of \citet{edwin1983} 
when the dependency of Equation~(\ref{mkt}) for the perturbed variables is used.
\\
Notice that Equation~(\ref{MHDWavesODE2}) and, a fortiori, Equation~(\ref{MHDWavesODE3}) are very reminiscent of Equation~(\ref{eq:pcomp}) in the Cartesian geometry. In the incompressible limit $\Gamma\to -k^2_z$ and the analogy is clear if one identifies $m/r$ with $k_y$ as done earlier. But more importantly, if one considers `thin tubes', the right hand side coefficient is dominated by $(m/r)^2$ (in the domain of interest, i.e. $r \approx R$), which is analogous to the limit of `nearly perpendicular propagation' considered in the Cartesian case. In particular we conclude again that Equation~(\ref{MHDWavesODE3}) thus becomes ignorant of and insensible to the value of the sound speed. Furthermore, the solutions decay away from the discontinuity surface, indicative of surface wave behaviour. Nodes are only found in the solutions at radial distances $r\gg R$, far away from the domain of interest.   

\subsubsection{Classic Alfv\'en waves}
\citet{edwin1983} concentrated on obtaining solutions to their Equation~(5), our Equation~(\ref{MHDWavesODE3}), for a piecewise uniform cylindrical plasma  and did not consider their Equation~(3b), our Equation~(\ref{MHDWavesODE4}), further.  Let us now focus on the solutions to Equation~(\ref{MHDWavesODE4}) and investigate what has happened to the classic Alfv\'{e}n waves when we move from an infinite uniform plasma to a cylindrical plasma with a piecewise constant density. 

First we consider the extreme case that $\rhoi = \rhoe$ so that the plasma is uniform. In that case $\omegaA^2$ is a constant and  we have a solution of the system of Equations~(\ref{MHDWavesODE1}) for any $m$, namely
\begin{equation}
\omega^2 = \omegaA^2  \label{AWUniformCylinder1}
\end{equation}
with
\begin{eqnarray}
 \xi_{\varphi} & \neq & 0, \qquad \xi_r   \neq
0, \qquad \xi_{z} = 0, \nonumber \\
P'& = & 0, \qquad \frac{{\rm d} P'}{{\rm d}r} = 0, \qquad \nabla \cdot \xii = 0.
\label{AWUniformCylinder2}
\end{eqnarray}
The only constraint is that the waves described by Equations~(\ref{AWUniformCylinder1})--(\ref{AWUniformCylinder2}) have to satisfy is $\nabla \cdot \xii = 0$. This can be done in many ways. The only restoring force is the magnetic tension force
${\bf T} = - \rho \omegaA^2 (\xi_r  \mathbf{1}_r + \xi_{\varphi}\mathbf{1}_{\varphi}) = - \rho \omegaA^2 \xii$.  Thus, in an infinite and uniform cylindrical plasma pure Alfv\'en waves are independent of the azimuthal number number $m$. 

We  turn to $\rhoi \neq \rhoe$. Again the solutions must safisfy the constraints given in Equations~(\ref{AWUniformCylinder1}) and (\ref{AWUniformCylinder2}). Since $\omega^2 = \omegaAi^2$ for $r \leq R$ and $\omega^2 = \omegaAe^2$ for $r > R$ we have Alfv\'en waves that live in the interior and in the exterior, respectively, of the flux tube. For $\omega^2 = \omegaAi^2$ the components $\xi_r$ and $\xi_\varphi$ are different from zero for $r \leq R$ and are identically zero for $r > R$. In addition $\xi_r = 0$ at the boundary $r=R$. In the particular case $m=0$, $\xi_r = 0$ everywhere. Conversely for $\omega^2 = \omegaAe^2$ the components $\xi_r$ and $\xi_\varphi$ are different from zero for $r > R$ and are identically zero for $r \leq R$. At the boundary $r=R$ again $\xi_r = 0$. Note that the frequencies are independent of the azimuthal wavenumber $m$. When we replace the piecewise constant density profile by a fully non-uniform density profile, the modes with $m=0$ are the only ones that survive as purely incompressible modes.

\subsubsection{Surface Alfv\'en waves}
\label{sec:surfa}

% Let us now see how  the solutions to Equation~(\ref{MHDWavesODE3}), for the fundamental radial modes with $m\neq 0$
% that have their phase velocities in the range $[\vai, \, \vae]$, change when we vary the density contrast between the internal and the external plasmas.  

\begin{figure}
	\centerline{\includegraphics[width=.4\columnwidth]{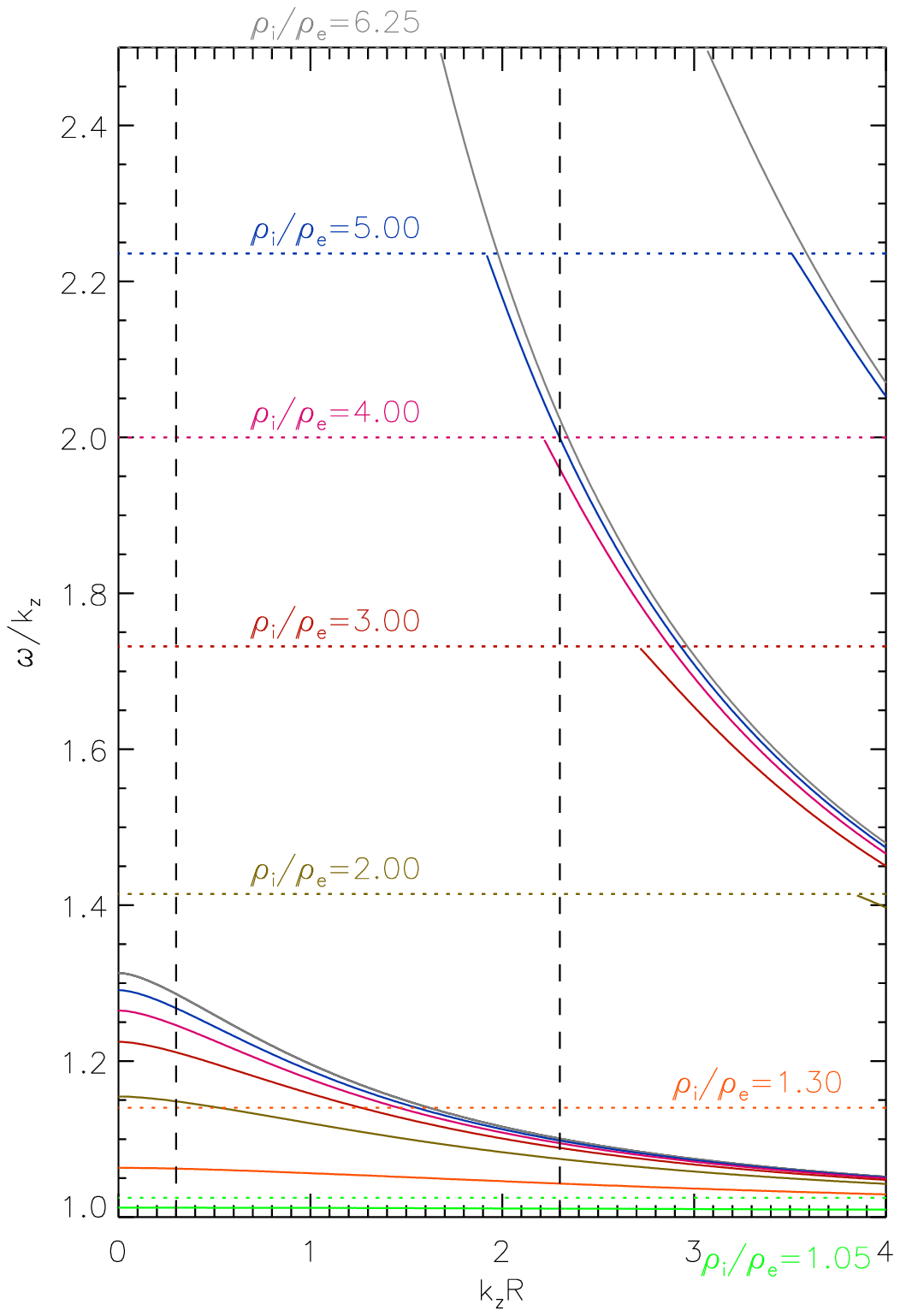}}
	\centerline{\includegraphics[width=.4\columnwidth]{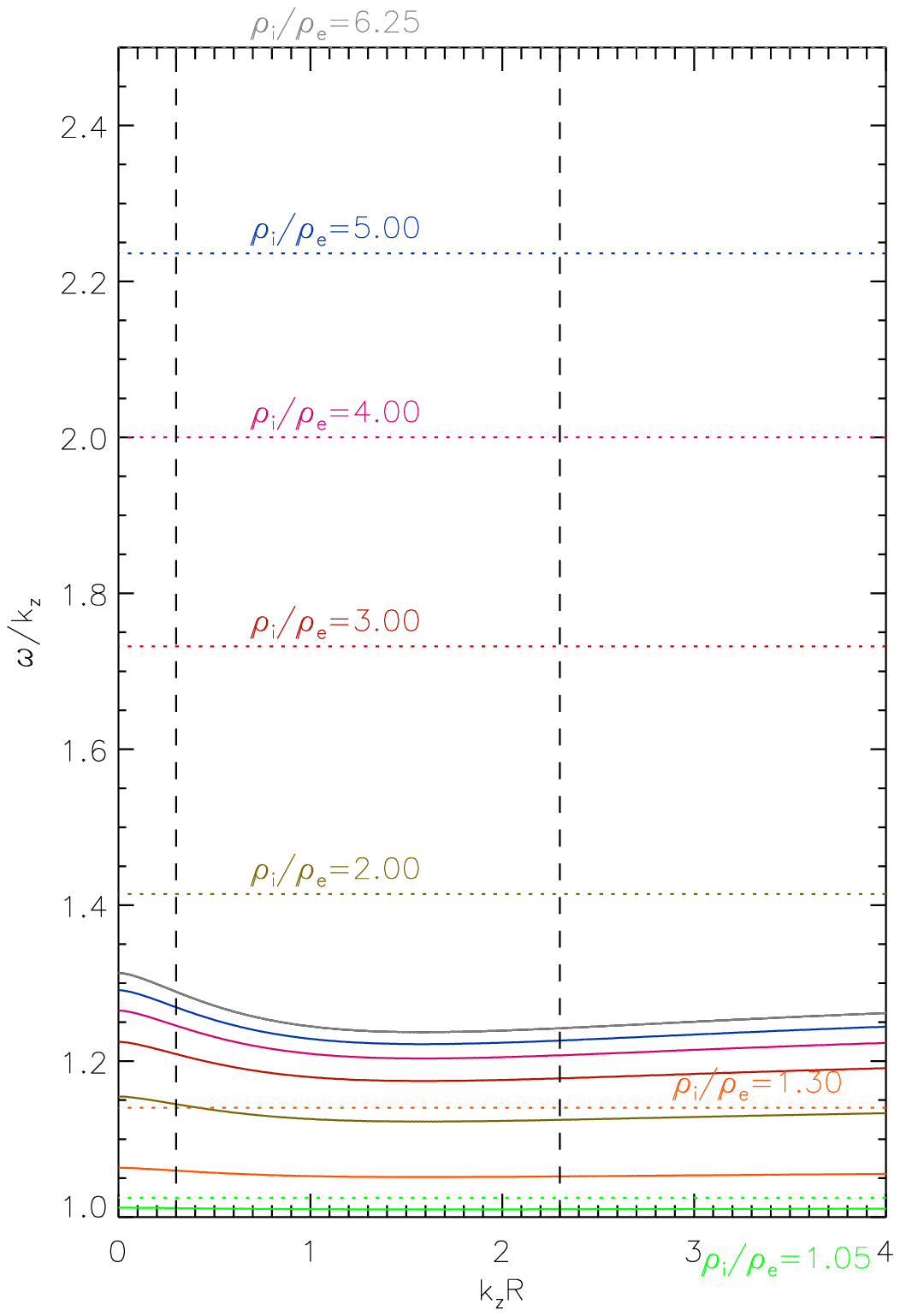}}
	\caption{Top graph: Dispersion diagram $\omega/k_z$ vs. $k_zR$ for compressible kink ($m=1$) modes with different density contrasts plotted with different colors. The respective external Alfv\'en velocity is displayed with a horizontal dashed line, also showing the density contrast. Bottom graph: Same as the top graph but for incompressible waves.}
	\label{fig:disp}
\end{figure}

Here we investigate the solutions to Equation~(\ref{MHDWavesODE3}) with $m\neq 0$. We vary $\rho_i - \rho_e$ and see what happens to the part of the spectrum with phase velocities between $\vai$ and $\vae]$. We start from the dispersion curves of solutions for $\rhoi - \rhoe = 1.5 \rhoe$, because this is analogous to Figure~4 in \citet{edwin1983}. We keep $ \rhoi$ constant and  decrease  $\rhoi - \rhoe$ from $1.5 \rhoe$ to 0 and follow the evolution of eigenmodes on the original dispersion curve. This evolution is illustrated in Figure~\ref{fig:disp} for $m=1$ modes, i.e., kink modes. 

In the top graph of Figure~\ref{fig:disp}, we show the dispersion diagram of compressible modes with the phase speed on the vertical axis and the normalized wave number on the horizontal axis, for different density contrasts in various colors. The associated external Alfv\'en speed $\vae$ (dashed line) and the density contrast are also indicated. In the solution, the internal Alfv\'en speed has always been normalized to $\vai=1$. In the bottom graph of Figure~\ref{fig:disp} we display the equivalent dispersion diagram but for incompressible modes. We can compare the top and bottom graphs of Figure~\ref{fig:disp} to assess the differences between compressible and incompressible waves. On purpose we have plotted the bottom graph on the same scale as the top graph in order to make the differences as clear as possible. The more striking difference is that the upper part of the bottom (incompressible) graph is empty. The dispersion curves in the top right belong to radial overtones of the kink modes. For radial overtones the total pressure perturbation $P'$ has an additional node in the internal region.  We notice the absence of the radial overtones in the incompressible dispersion diagram. The dispersion curves in the bottom of both graphs in Figure~\ref{fig:disp} belong to the fundamental radial kink mode.  The fundamental radial modes survive in the incompressible limit while radial harmonics are absent. Radial harmonics need compression to exist. In contrast the fundamental radial modes do not need compression to exist.  Compression is a typical characteristic of magneto-sonic waves.  Hence the fundamental radial modes do not have the typical properties of fast magneto-sonic modes. Instead, they behave like surface Alfv\'en waves. 

\begin{figure}
	\centerline{\includegraphics[width=.9\columnwidth]{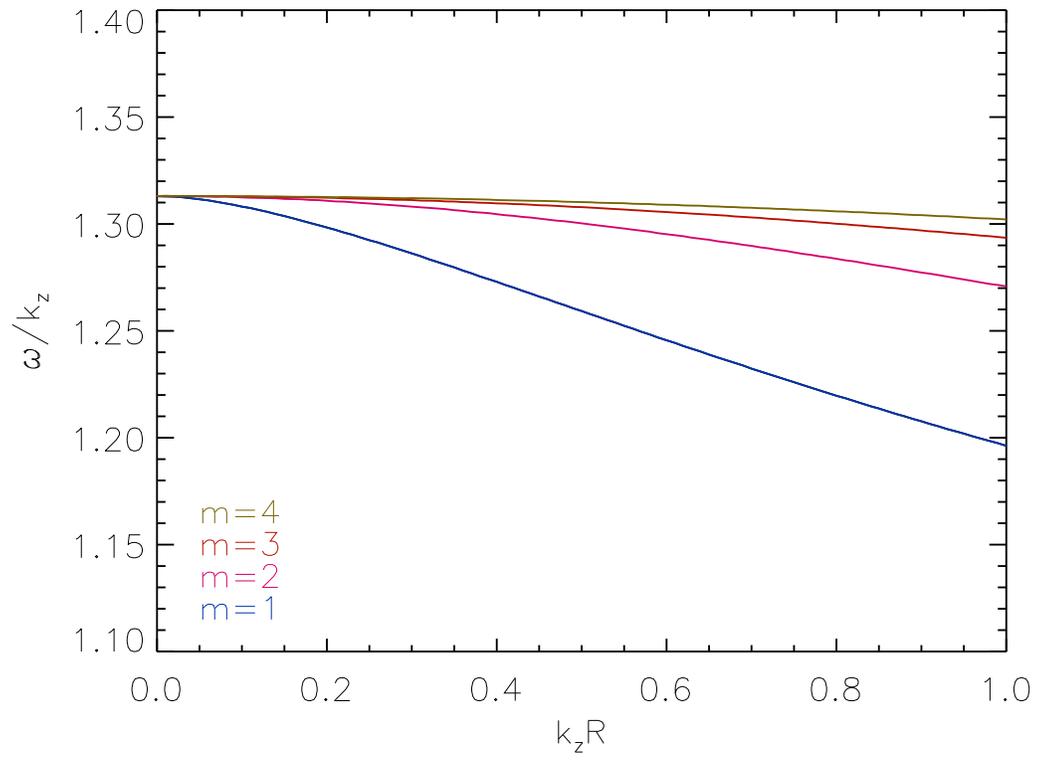}}
	\caption{Dispersion diagram $\omega/k_z$ vs. $k_zR$ for the fundamental radial modes with $\rhoi/\rhoe = 6.25$ and different values of $m$.}
	\label{fig:dispm}
\end{figure}

The behavior of the fundamental radial modes in the thin tube limit ($k_z R \ll 1$) is the same in both compressible and incompressible cases, i.e., their phase velocity tends to the kink velocity, $\vk$, namely
\begin{equation}
 \vk = \sqrt{\frac{\rhoi \vai^2 + \rhoe \vae^2}{\rhoi+\rhoe}}. \label{TTappspeed}
\end{equation}
The behavior of compressible and incompressible waves is slightly different for large $k_z R$, i.e., the phase velocity of the compressible waves tends to $\vai$ while that of incompressible waves remains between $\vai$ and the kink velocity, $\vk$. We display in Figure~\ref{fig:dispm} the dispersion diagram of the fundamental radial modes with different values of $m$ and a fixed value of the density contrast. In the thin tube limit, i.e., $k_z R \ll 1$, the results for the different values of $m$ overlap and their phase velocity tends to the kink velocity, $\vk$. The differences between the modes with different values of $m$ grow when we take larger values of $k_zR$. For $k_z R \ll 1$ the frequencies are to a good approximation independent of $k_z R$ and independent of $m$. The behavior of the fundamental radial modes is reminiscent of Alfv\'en waves in uniform plasmas of infinite extent and of surface Alfv\'en waves in a true discontinuity. 

\begin{figure}
	\centerline{\includegraphics[width=.8\columnwidth]{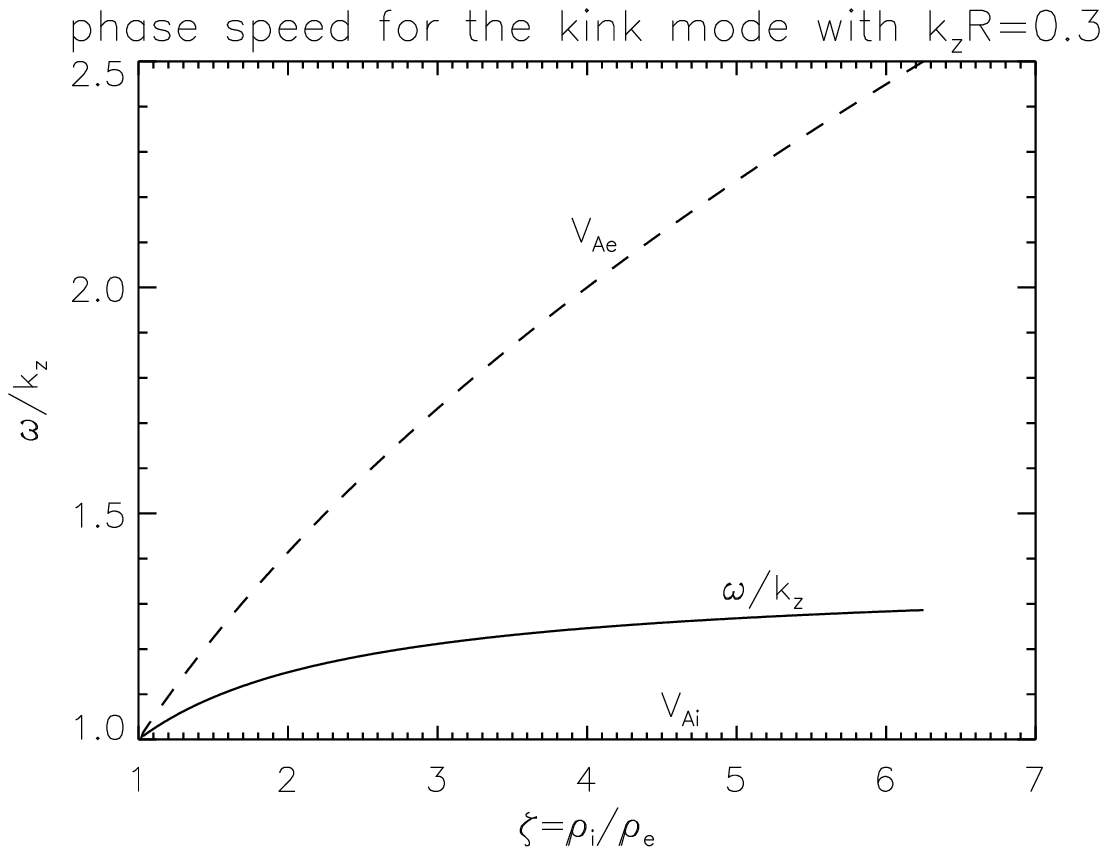}}
	\centerline{\includegraphics[width=.8\columnwidth]{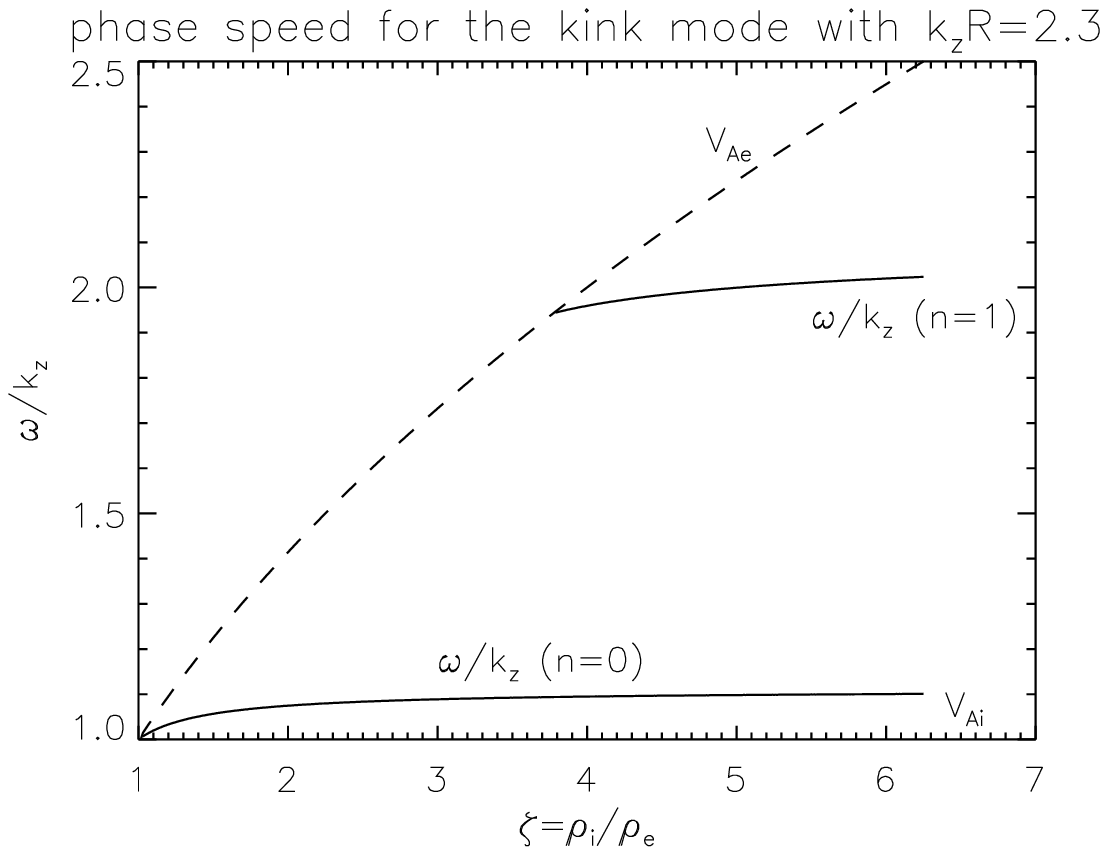}}
	\caption{Evolution of the phase speed as a function of density contrast, for $k_zR=0.3$ (top graph) and $k_zR=2.3$ (bottom graph). The external Alfv\'en speed is indicated with a dashed line, while the internal Alfv\'en speed is set to 1. These two graphs are vertical cuts in the top diagram of Figure~\ref{fig:disp}, along the indicated vertical long dashed lines. In the bottom graph $n=0$ means the fundamental radial mode and $n=1$ means the first radial overtone.}
	\label{fig:disp2}
\end{figure}

In Figure~\ref{fig:disp2} the evolution of the phase speed with varying density contrast is shown for two different wave numbers, namely $k_zR = 0.3$ and $k_zR = 2.3$ (indicated as vertical dashed lines in the top graph of Fig.~\ref{fig:disp}). Several observations can be made from the results of Figure~\ref{fig:disp2}. It is clear that the dispersion curve for the fundamental radial kink mode is always between the external and internal Alfv\'en speed. When these two values approach each other, i.e., $\rhoi-\rhoe \to 0$, the dispersion curve eventually collapses to the Alfv\'en velocity, $\omega/k_z=\vai$. In that sense the fundamental radial mode is the descendant of the Alfv\'en mode of the uniform case. In fact, we have calculated that all curves that start at the kink speed (Equation~(\ref{TTappspeed})) collapse to the internal Alfv\'en speed as the density difference $\rhoi-\rhoe$ goes to 0. The evolution of the radial overtones is entirely different from that of the fundamental radial mode. As the density contrast decreases, the radial overtones are less well confined to the magnetic cylinder.  When the density contrast reaches a critical point, confinement is completely breached and energy leaks away in the form of MHD radiation. Then, radial overtones become leaky modes with complex frequencies because of damping due to MHD radiation \citep[see, e.g.,][]{wilson1981,spruit1982,cally1986,goossens1993}. When the density contrast decreases, the phase speeds of these radial overtones do not collapse to the Alfv\'en velocity. These modes are not related to the Alfv\'en mode of the uniform case, but rather to fast modes. As such, they become leaky when the density contrast is too low. This same argument explains the peculiar behavior of the fundamental radial mode with $m=0$, i.e., the sausage mode (not displayed in Figure~\ref{fig:disp}).

\subsubsection{Vorticity}
Here we focus on the role of vorticity. As explained before, vorticity is a typical characteristic of Alfv\'en waves. Let us now consider the  vorticity of the solutions of Equation~(\ref{MHDWavesODE3}). 

The reader of the original paper of \citet{edwin1983} might have the impression that vorticity for the solutions of  a piecewise constant equilibrium vanishes.  However, visual inspection of the radial variation of $\xi_{ \varphi}$ shows that it is discontinuous at $r=R$ with opposite values to the left and right of the boundary \citep[see Figure~1b of][corresponding to the $m=1$ mode]{goossens2009}. Actually \citet{terradas2008nonlinear} studied the Kelvin-Helmholtz instability triggered by the velocity shear in  $\xi_{ \varphi}$ at the boundary. So there is vorticity present in this configuration.  For mathematical simplicity let us adopt the thin tube (TT) approximation, i.e., $k_z R \ll 1$, so that the dispersion relation for modes with $m\neq0$ is \citep[see details in, e.g.,][]{goossens1992,goossens2009}
\begin{equation}
\omega^2 = \frac{ \rhoi \omegaAi^2 + \rhoe \omegaAe^2}{\rhoi + \rhoe} \equiv \omega_{\rm k}^2.
\label{TTapp}
\end{equation}
Note that this is exactly the same dispersion relation as for incompressible surface Alfv\'en waves at a true discontinuity (Equation~(\ref{eq:kinkfreq})). Equation~(\ref{TTapp}) is independent of $m$.

% In the TT approximation, $\Gamma (\omega^2) \to 0$ in Equation~(\ref{MHDWavesODE3}) and the perturbation of the total pressure, $P'$, for the fundamental radial modes can be written as
% \begin{equation}
% P' = \left\{  \begin{array}{lll}
% P'(R) \left( \frac{r}{R} \right)^m, & \textrm{if}, & r \leq R, \\
% P'(R) \left( \frac{R}{r} \right)^m, & \textrm{if}, & r > R,
% \end{array} \right. \label{eq:presTT}
% \end{equation}
% with $P'(R)$ the perturbation of the total pressure at $r=R$. From Equation~(\ref{TTapp}) it follows that 
% \begin{eqnarray}
%  \rhoi (\omega^2 - \omegaAi^2) &=& \frac{ \rhoi \rhoe}{ \rhoi + \rhoe} (\omegaAe^2 - \omegaAi^2 ) > 0, \nonumber \\
% \rhoe (\omega^2 - \omegaAe^2)  &=& - \rhoi (\omega^2 - \omegaAi^2), 
% \end{eqnarray}

%  and
% \begin{equation}
%  \frac{\rm d}{{\rm d}r} \left[ \rho (\omega^2 - \omegaA^2) \right] = 
% - \frac{2}{R} \frac{\rhoi \rhoe}{\rhoi + \rhoe} (\omegaAe^2 - \omegaAi^2 ) \delta(r -R). \label{eq:delta}
% \end{equation}
% Here $H(r-R)$ is the Heaviside step function and $\delta$ is the delta function.  

From Equation~(\ref{TTapp}) it follows that 
\begin{equation}
 \frac{1}{\rho (\omega^2 - \omegaA^2)} = \frac{\rhoi + \rhoe}{\rhoi \rhoe} \frac{1}{\omegaAe^2 - \omegaAi^2} \left[ 1 - 2 H(r-R) \right]. \label{eq:heviside2}
\end{equation}
Here $H$ is the Heaviside step function as defined in Equation~(\ref{eq:heviside}). Since $P'$ is continuous at $r= R$ it follows from Equations~(\ref{Vorticity2}) and (\ref{eq:heviside2}) that 
\begin{equation} 
(\nabla \times \xii) \cdot \mathbf{1}_z  = -2 i \frac{m}{R} P' \frac{\rhoi + \rhoe}{\rhoi \rhoe} \frac{1}{\omegaAe^2 - \omegaAi^2} \delta (r-R),
\label{Vorticity3}
\end{equation}
where $\delta$ is again the Dirac delta function. Thus, the solutions to Equation~(\ref{MHDWavesODE3}) for the fundamental radial mode of MHD waves with frequencies  $\omega \in [\omegaAi, \,\omegaAe] $ have vorticity but it is concentrated as a delta function at the boundary. This is exactly the same behavior obtained for surface Alfv\'en waves at a true discontinuity (Sect.~\ref{sec:discsurf}). This is a rather pathological situation that finds its origin in the fact that the equilibrium has been forced to be piecewise uniform. The singularities of the continuous spectrum are all concentrated in the point $r=R$. Note that this is true for all modes with $m\neq 0$. This result is a strong argument in favor of a classification of the fundamental radial modes as surface Alfv\'en waves instead of fast body modes.

\subsection{Continuous density variation}
In the previous Subsections we have studied the properties of the waves for a piecewise constant density profile. Here we replace the discontinuity in density with a continuous variation of density in an intermediate layer of thickness $l$. Thus, density is non-uniform in the interval $]R - l/2, \,R +l/2[$. Since $\omegaA(r)$ is non-constant in the interval $]R - l/2, \,R +l/2[$, it follows from Equation~(\ref{Vorticity2}) that $(\nabla \times \xii) \cdot \mathbf{1}_z \neq 0$ in that interval. This means that vorticity is spread out over the interval with non-uniform density.

At this point it is instructive to note that when we replace the piecewise constant density of \citet{edwin1983} by a continuous variation of density all wave modes with phase velocities between $\vai$ and $\vae$ are in the Alfv\'en continuous spectrum. As a consequence, the waves with $m\neq0$ undergo resonant damping. The fundamental conservation law for resonant Alfv\'en waves was obtained by \citet{sakurai1991} in ideal MHD and by \citet{goossens1995} in dissipative MHD for the driven problem, and by \citet{tirry1996} for the eigenvalue problem. For a straight magnetic field the conserved quantity is the total pressure perturbation, $P'$.

We denote by $\ra$ the position of the Alfv\'en resonant point where $\omega = \omegaA(\ra)$ and use a Taylor expansion of $\omega^2 - \omegaA^2$ in the vicinity of $\ra$, namely
\begin{equation}
 \omega^2 - \omegaA^2 \approx s \Delta_{\rm A} + \mathcal{O}\left(s^2\right)
\end{equation}
 where $\Delta_{\rm A} = \frac{\rm d}{{\rm d}r} \left( \omega^2 - \omegaA^2  \right)$ and $s=r-\ra$. It follows from the third equation in Equations~(\ref{MHDWavesODE1}) that in ideal MHD $\xi_\varphi$ diverges as $1/s$ near the resonant point. Equation~(\ref{Vorticity2}) implies that the singular behavior of $\left( \nabla \times {\bf \xi} \right) \cdot {\bf 1}_z$ is $1/s^2$ and hence stronger than that of $\xi_\varphi$. Hence vorticity is different from zero everywhere in the non-uniform plasma when ${{\rm d} \omegaA^2}/{{\rm d}r} \neq 0$, but it is by far largest at the resonant position $r=\ra$. 

Now we can compare the behavior of vorticity in the non-uniform case with that in the particular case of a piecewise constant density profile (Equation~(\ref{Vorticity3})). The delta function behavior at $r=R$ obtained for the piecewise constant profile is a very pathological and peculiar situation. In the non-uniform case vorticity is spread everywhere in the region of non-uniform density.

Conversely for a wave with its frequency in the slow continuum, $\xi_z$ and $\nabla \cdot {\bf \xi}$ are singular at the slow resonant position $\rc$ where  $\omega = \omegaC(\rc)$. Both quantities diverge as $1/s$ with $s$ now defined as $s=r-\rc$. For coronal conditions the Alfv\'en continuum and the slow continuum do not overlap. Hence when we study MHD waves with frequencies in the Alfv\'en continuum we do not have to worry about the slow continuum. The situation is different in, e.g., thin threads of prominences \citep[see][]{soler2009}, and photopheric flux tubes in which the frequency of the radially fundamental modes with $m\neq 0$ is in both the Alfv\'en and slow continua.

\begin{figure}
	\centerline{\includegraphics[width=.8\columnwidth]{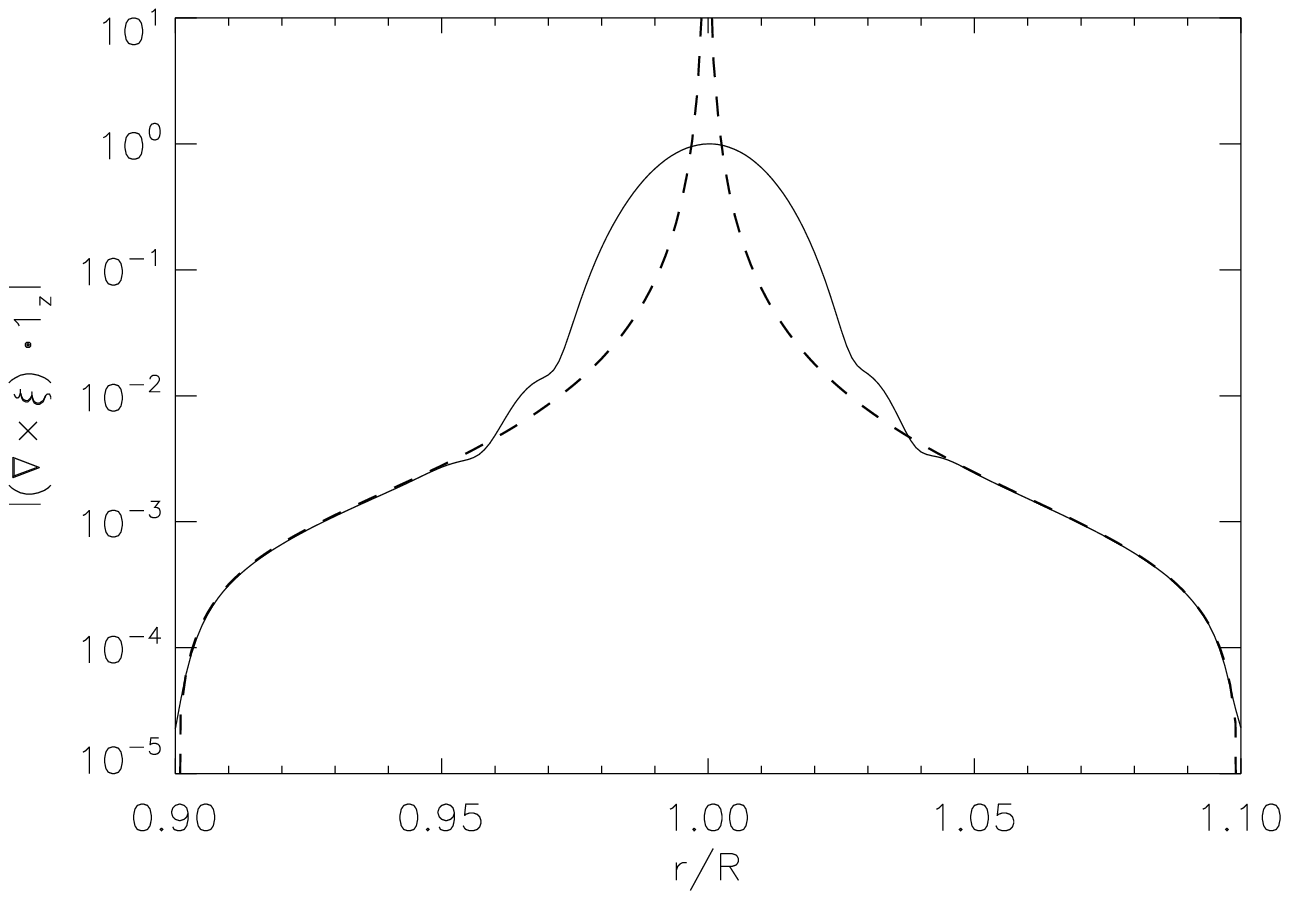}}
	\centerline{\includegraphics[width=.8\columnwidth]{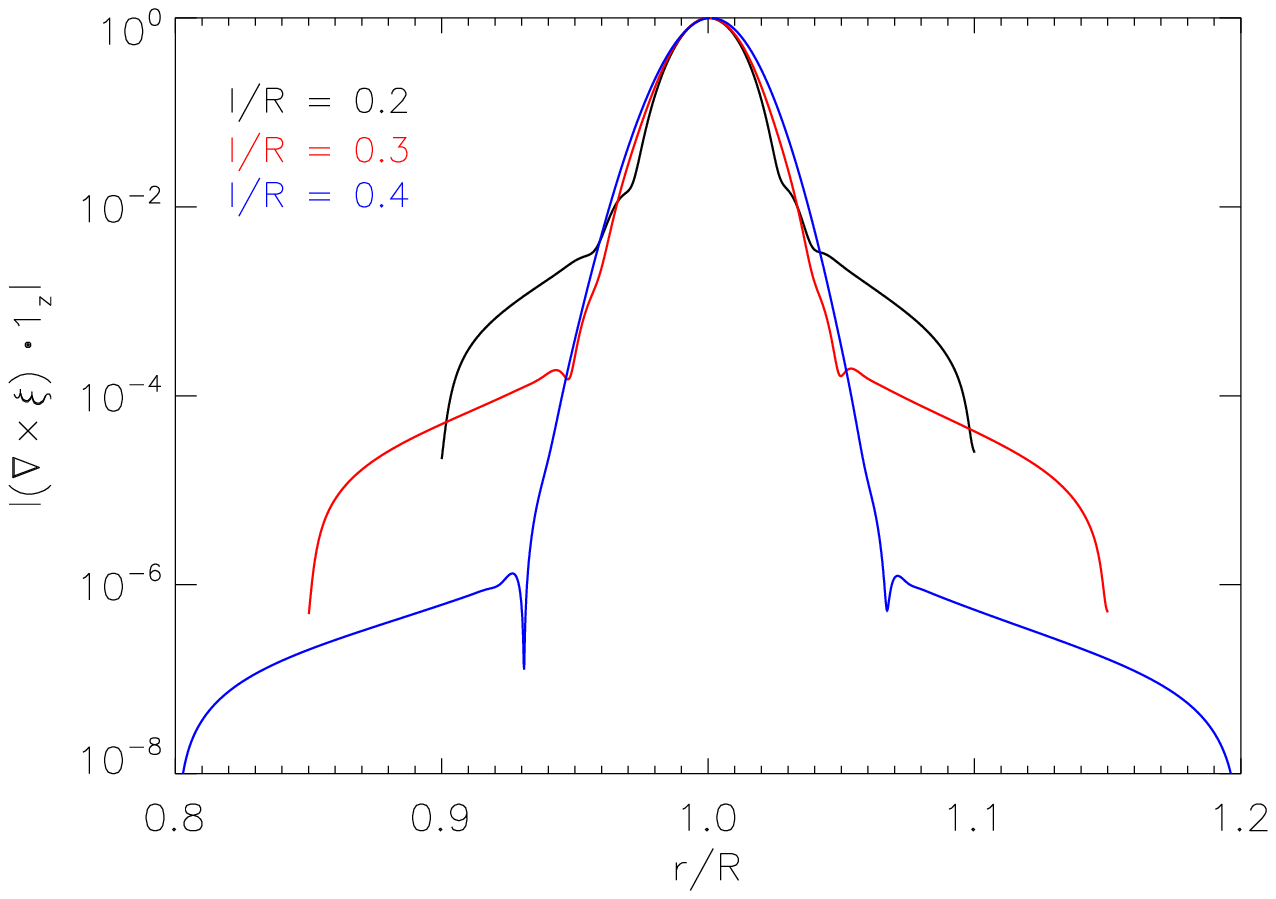}}
	\caption{Top: Absolute value of the $z$-component of vorticity (in dimensionless units) in the non-uniform layer for the fundamental radial $m=1$ mode with $k_z R = 0.1$, $\rhoi/\rhoe = 2$, and $l/R = 0.2$. The solid line is the resistive result with $R_{\rm m} = 10^7$. The dashed line shows the ideal spread of vorticity due to non-uniformity computed from Equation~(\ref{Vorticity2}) using the numerically obtained $P'$. Bottom: Same as the top panel for the resistive result but with different values of $l/R$.}
	\label{fig:nonuniform}
\end{figure}

Next we compute the eigenfrequencies and the perturbations of the fundamental radial mode numerically. In order to have non-singular eigensolutions we need to remove the singularity by including dissipative effects. For that reason we compute eigenmodes of non-uniform equilibrium states in resistive MHD.  We add the term $\eta \nabla^2 {\bf B}'$ to the right-hand side of the linearized induction equation (the second equation of Equations~(\ref{basiceq})), where $\eta$ is the coefficient of magnetic diffusion or resistivity. For simplicity we take $\eta$ as a constant. We define the magnetic Reynolds number as $R_{\rm m} = \vai R/\eta$.

Our numerical procedure uses the PDE2D code \citep{sewell} based on finite elements to solve the eigenvalue problem defined by Equations~(\ref{basiceq}) in our equilibrium. The numerical integration of Equations~(\ref{basiceq}) is performed from the cylinder axis, $r=0$, to the finite edge of the numerical domain, $r=r_{\rm max}$, which is located far enough to obtain a good convergence of the solution and to avoid numerical errors. This means that we take $r_{\rm max} \gg R$. We use a nonuniform grid with a large density of grid points within the nonuniform interval $]R - l/2, \,R +l/2[$. The nonuniform grid also allows us to correctly describe the small spatial scales of the eigenfunctions within the nonuniform region due to the Alfv\'en resonance. The PDE2D code uses a collocation method and the generalized matrix eigenvalue problem is solved using the shifted inverse power method. The output of the program is the closest complex eigenvalue to an initial provided guess and the corresponding perturbations.

%{\bf Some of Inigo's computations should be put here. Maybe one Figure with Re and Im of vorticity  and some text explaining it? To be discussed}

First we use the components of the displacement, obtained numerically, to compute $(\nabla \times \xii) \cdot \mathbf{1}_z$ in the non-uniform region for $l/R = 0.2$ (see Figure~\ref{fig:nonuniform}a). The remaining parameters are given in the caption of the Figure. Since $(\nabla \times \xii) \cdot \mathbf{1}_z$ is a complex quantity, we plot its absolute value. We use dimensionless units so that the maximum of $\left| (\nabla \times \xii) \cdot \mathbf{1}_z \right|$ has been set to unity. Vorticity is present in the whole non-uniform region and is maximal at the Alfv\'en resonance position, $\ra \approx R$. We overplot in  Figure~\ref{fig:nonuniform}a the ideal $\left| (\nabla \times \xii) \cdot \mathbf{1}_z \right|$ computed from Equation~(\ref{Vorticity2}) using the obtained $P'$ from the numerical code (see the dashed line). From either curve we conclude that non-uniformity spreads out vorticity but the resonant behavior is so strongly present that the values of $(\nabla \times \xii) \cdot \mathbf{1}_z$ close to the ideal singularity totally overpower the values away from that position.

Equation~(\ref{Vorticity2}) gives us the ideal behavior of vorticity and doesn't include the effect of diffusion. Diffusion removes the singular $1/s^2$ behavior of  $(\nabla \times \xii) \cdot \mathbf{1}_z$ found in ideal MHD and limit $(\nabla \times \xii) \cdot \mathbf{1}_z$ to finite values. The remnant of the ideal $1/s^2$ behavior is, however, still clearly present. In fact, the effect of diffusion is important in a dissipative layer of width $\delta_{\rm A}$ around the resonance position given by \citep[see, e.g.,][]{sakurai1991}
\begin{equation}
 \delta_{\rm A} = \left( \frac{\omega}{\left| \Delta_{\rm A} \right|} \eta \right)^{1/3}. \label{eq:da}
\end{equation}
For the particular case of Figure~\ref{fig:nonuniform}a, the dissipative layer extends approximately in the interval $0.95 \lesssim r/R \lesssim 1.05$. The resistive result of  Figure~\ref{fig:nonuniform}a corresponds to $R_{\rm m} = 10^7$, while the actual Reynolds number in the corona is believed to be around $R_{\rm m} = 10^{12}$. Using the actual value of $R_{\rm m}$ is unpractical from the computational point of view as it requires taking an enormous number of grid points in the numerical domain. We therefore use a smaller value of $R_m$, and consequently a smaller number of grid points, but the qualitative effect of magnetic diffusion remains correctly described. Thus, the width of the dissipative layer around the resonance position would be extremely thin if the actual Reynolds number would be used. We have made sure that the Reynolds numbers used in the computations are in the so-called {\sl plateau} where the damping rate by resonant absorption is independent of $R_m$ and so the wave behavior is not dominated by diffusion \citep[see, e.g.,][]{poedts1991,vd2004}.

Now we vary the thickness of the non-uniform region, $l$. Figure~\ref{fig:nonuniform}b displays vorticity in the nonuniform layer for different values of $l/R$. For comparison purposes we have set $\max\left| (\nabla \times \xii) \cdot \mathbf{1}_z \right| = 1$ in all cases. As before, vorticity is larger near the resonance position, but vorticity spreads along the whole region of non-uniform density. In a fully non-uniform equilibrium, vorticity would spread out over the whole domain.

% Now we vary the thickness of the non-uniform region, $l$.  Since $\Delta_{\rm A}$ depends on the derivative of the Alfv\'en frequency at the resonance position, $\delta_{\rm A}$ changes when $l$ varies. In order to have the influence of diffusion confined to a region of fixed thickness, we keep $\delta_{\rm A}$ approximately constant in the computations when we vary $l$. We do so by selecting the value of $\eta$, and so $R_{\rm m}$, accordingly. Figure~\ref{fig:nonuniform}b displays vorticity in the nonuniform layer for different values of $l/R$. For comparison purposes we have set $\max\left| (\nabla \times \xii) \cdot \mathbf{1}_z \right| = 1$ in all cases. As before, vorticity is larger near the resonance position, but vorticity spreads along the whole region of non-uniform density. Outside the dissipative layer, vorticity decreases as $l/R$ is increased. This can be easily understood looking at Equation~(\ref{Vorticity2}). As $l/R$ increases, the derivative  of the Alfv\'en frequency decreases, causing the decrease of vorticity.

\section{Discussion}
\label{sec:conclusions}
In this paper we have studied the different properties of linear Alfv\'en waves and magneto-sonic waves in uniform and non-uniform plasmas. First in a uniform plasma of infinite extent, we have reiterated that Alfv\'{e}n waves are driven solely by the magnetic tension force and that they are the only waves that propagate vorticity. The displacements are vortical and incompressible. On the contrary, magneto-sonic waves are driven by both the total pressure force and the magnetic tension force. The displacements are compressible and have no vorticity.

Then we have moved to non-uniform plasmas and have investigated how MHD waves are modified by non-uniformity. For the case of a true discontinuity in the Alfv\'en velocity we find that the incompressible surface Alfv\'en waves have vorticity equal to zero everywhere except at the discontinuity, where all vorticity is concentrated. The behavior of the surface Alfv\'en waves is clearly different from that of the classic Alfv\'en waves in a uniform plasma of infinite extent, which have vorticity different from zero everywhere.

Subsequently we have considered the case of MHD waves superimposed on a 1D non-uniform straight cylinder with constant magnetic field.  For the particular case of a piecewise constant density profile as studied by \citet{edwin1983}, we find that the fundamental radial modes of the non-axisymmetric ($m\neq0$) waves with phase velocity between $\vai$ and $\vae$ have properties remarkably similar to those of surface Alfv\'en waves in a true discontinuity. In this pathological situation vorticity is present as a delta function at the cylinder boundary. When the discontinuity in density is replaced with a continuous variation of density, vorticity is spread out over the whole interval with non-uniform density. The fundamental radial modes of the non-axisymmetric waves do not need compression to exist unlike the radial overtones.

With these insights we may now also interpret the physics behind the computational results presented in \citet{vd2007}. In that article the evolution of the $m=1$ kink mode frequency was followed while the thickness of the inhomogeneous layer around the flux tube was increased ($l/R \nearrow$). It was found that the kink mode frequency joined the Alfv\'en continuum when $l/R$ passed a critical threshold. Indeed, understanding these waves as surface Alfv\'en waves, now lets us conclude that the frequencies return to the Alfv\'en continuum as the surface `goes away'. No surface, no surface mode. See also \citet{sedlacek1971a} for a similar interpretation.

We would like to stress that the importance of the labels Alfv\'en or fast is not in the names themselves but in the properties that are intrinsically associated with these names. Due to plasma non-uniformity MHD waves have mixed properties and cannot be classified as pure Alfv\'{e}n or pure magneto-sonic waves. However, there are basic characteristics that remain strongly related to the wave type.  Our results show that in 1D magnetic cylinders the fundamental radial modes of the waves with $m\neq0$ and phase velocities between $\vai$ and $\vae$ have not the typical properties expected for fast magneto-sonic waves. Instead, their properties resemble very much those of surface Alfv\'{e}n waves in a true discontinuity. For this reason we call these waves surface Alfv\'{e}n waves as was already done by \citet{wentzel1979}. 

Here we go back to the discussion on the nature of the ubiquitous, transverse waves as observed in the solar corona \citep{tomczyk2007,depontieu2007,mcintosh2011}. In view of the results of the paper by \citet{goossens2009} and the present paper, the controversy about the interpretation of the observed waves is lifted. The fundamental radial modes of kink ($m=1$) waves with phase velocity between the internal and external Alfv\'en velocities can be considered as surface Alfv\'en waves (or Alfv\'enic waves in the nomenclature of \citealt{goossens2009}). The two interpretations refer to the same physical phenomenon of a wave dominated by tension forces. The controversy was also partly due to the claim by \citet{vd2008} that Alfv\'en waves have to be torsional, i.e., axi-symmetric with azimuthal wavenumber $m=0$. Axi-symmetric MHD waves do not displace the axis of the magnetic cylinder and the cylinder as a whole. The view that Alfv\'en waves need to be axi-symmetric is too narrow. Anyway, in a non-uniform plasma there is a continuous spectrum of Alfv\'en waves with frequencies independent of the azimuthal wavenumber, $m$.

Note that the observations of \citet{tomczyk2007} are not the first observations of Alfv\'enic waves, but that they are the first to observe the ubiquity of these waves. Surface Alfv\'en waves as described in the present paper have been observed on earlier accounts although the authors at that time did not realize that they had indeed observed Alfv\'enic waves. \citet{goossens2009} pointed out that accepting resonant absorption as damping mechanism of the transverse MHD waves observed with the Transition-Region And Coronal Explorer (TRACE) implied that these waves are surface Alfv\'en waves or, in the nomenclature of \citet{goossens2009}, Alfv\'enic waves. The fundamental radial modes of kink waves with phase velocity between the internal and external Alfv\'en velocities are surface Alfv\'en waves. Hence the TRACE observations of transverse MHD waves starting in 1999 with \citet{schrijver1999}, \citet{aschwanden1999}, and \citet{nakariakov1999} were observations of surface Alfv\'en waves.

% {\bf Tom's addition (to be discussed): We would like to stress though, that the description in terms of kink waves in cylinders offers a more detailed model for the seismology and calculation of the energy budget of the observed waves (of course, only if you believe that magnetic cylinders are an adequate description of coronal loop structures). }
% 

% We would like to stress that the importance of the labels Alfv\'en or fast is not in the names themselves but in the properties that are intrinsically associated with these names. For example, in uniform plasmas fast magneto-sonic waves are compressible waves, whereas  Alfv\'en waves are incompressible. In non-uniform plasmas MHD waves have mixed properties so that, in general, there are no pure Alfv\'{e}n or magneto-sonic waves. However, there are some properties that remain strongly related to the wave type. Thus in non-uniform plasmas surface Alfv\'en waves are almost incompressible, whereas fast waves still need compression to exist. To label the almost incompressible waves observed in the solar corona as fast waves is in clear contradiction with the intrisic properties of fast waves. However, the name Alfv\'en waves is justified. 

We finally note that although the observed waves can be interpreted as a type of Alfv\'en waves it is crucial to consider an adequate description of the coronal loop structures in which these waves propagate. Hence the use of expressions for Alfv\'en waves in uniform plasmas of infinite extent may be inadequate for the study of waves propagating in the solar corona. A description in terms of surface Alfv\'en kink ($m=1$) waves in cylinders may offer a more detailed model for the seismology and calculation of the energy budget of the observed waves.

\acknowledgements{
MG and JA acknowledge support from KU Leuven via GOA/2009-009.
\\
MG, RS, IA, and JT acknowledge the support from the Spanish MICINN and FEDER funds through project AYA2011-22846. 
\\
JA acknowledges support by a Marie Curie International Outgoing Fellowship under the EU's 7th Framework Programme with grant number PIOF-GA-2008-219943.
\\
RS also acknowledges support from a Marie Curie Intra-European Fellowship within the European Commission 7th Framework Program  (PIEF-GA-2010-274716).
\\
RS and JT acknowledge support from CAIB through the `grups competitius' scheme.
\\
TVD acknowledges funding from the Odysseus Programme of the FWO Vlaanderen and from the EU's 7th Framework Programme as an ERG with grant number 276808.
\\
IA acknowledges support by a Ram\'on y Cajal Fellowship by the Spanish Ministry of Economy and Competitiveness (MINECO).
\\
JT also acknowledges support from the Spanish Ministerio de Educaci\'on y Ciencia through a Ram\'on y Cajal grant.}

\bibliographystyle{apj} 
\bibliography{refs}

\end{document}